\documentclass[journal,10pt]{IEEEtran}
\usepackage{cite}
\usepackage{amsmath,amssymb,amsfonts}
\usepackage{algorithmic}
\usepackage{lipsum,adjustbox}
\usepackage{verbatim}
\usepackage{graphics}
\usepackage{stackengine}
\usepackage{pgfplots}
\pgfplotsset{width=7cm,compat=1.8}
\usepackage{multirow}
\usepackage{soul}
\usepackage{amssymb}
\usepackage{amsmath}
\usepackage{algorithm}
\usepackage{algorithmic}
\usepackage{graphicx}
\usepackage{subcaption}
\usepackage{textcomp}
\usepackage{booktabs}
\usepackage{stfloats}
\usepackage{scalefnt}
\usepackage{mathrsfs}       
\usepackage[nolist]{acronym}
    
 \usepackage{bm}

\newcommand{\figref}[1]{Figure~\ref{#1}}

\newcommand{\tabref}[1]{Table~\ref{#1}}

\newcommand{\LeftP} {\left\lbrace }
\newcommand{\RightP}{\right\rbrace }

\newcommand{\compl}{\mathbb{C}}         

\newcommand{\ma}  [1]{ \bm{#1} } 

\newcommand{\Ex}[1]{\mathrm{E}\left[ #1\right]} 
\newcommand{\trace}[1] {\mathrm{trace}\LeftP #1 \RightP }

\newcommand{\Vect}  [1] {\mathrm{vec}  \LeftP #1 \RightP } 

\newcommand{\set} [1]{{\mathcal {#1}}} 
\newcommand{\Kon} {\set{K}_{\text{on}}} 
\newcommand{\Kp} {\set{K}_{\text{p}}} 
\newcommand{\Kd} {\set{K}_{\text{d}}} 

\newcommand{\RS} {\set{RS}} 
\newcommand{\URS} {\set{URS}} 

\newcommand{\Kf} {\set{K}_{f}} 
\newcommand{\Kt} {\set{K}_{t}} 

\begin{acronym}
\acro{DSRC}{dedicated short-range communications}
\acro{C-ITS}{cooperative intelligent transport system}
\acro{RSU}{road side unit}
\acro{TDL}{tapped delay line}
\acro{ITS}{Intelligent Transportation Systems}
\acro{IEEE}{Institute of Electrical and Electronics Engineers}
\acro{WAVE}{Wireless Access in Vehicular Environment}
\acro{V2V}{vehicle-to-vehicle}
\acro{V2I}{vehicle-to-infrastructure}
\acro{CCH}{control channel}
\acro{SCH}{service channels}
\acro{STS}{short training symbols}
\acro{LTS}{long training symbols}
\acro{SS}{signal symbol}
\acro{SoA}{state-of-the-art}
\acro{DPA}{data-pilot aided}
\acro{STA}{spectral temporal averaging}
\acro{CDP}{constructed data pilots}
\acro{TRFI}{time domain reliable test frequency domain interpolation}
\acro{MMSE-VP}{minimum mean square error using virtual pilots}
\acro{iCDP}{Improved CDP}
\acro{SBS}{symbol-by-symbol}
\acro{FBF}{frame-by-frame}
\acro{E-TRFI}{Enhanced TRFI}
\acro{SR-CNN}{super resolution CNN}
\acro{DN-CNN}{denoising CNN}
\acro{RBF}{radial basis function}
\acro{CNN}{convolutional neural network}
\acro{TS-ChannelNet}{Temporal spectral ChannelNet}
\acro{WSSUS}{wide-sense stationary uncorrelated scattering}
\acro{TDR}{transmission data rate}
\acro{LSTM}{long short-term memory}
\acro{ALS}{accurate LS}
\acro{SLS}{simple LS}
\acro{ChannelNet}{channel network}
\acro{ADD-TT}{average decision-directed with time truncation}
\acro{WI}{weighted interpolation}
\acro{DD}{decision-directed}
\acro{SR-ConvLSTM}{super resolution convolutional long short-term memory}
\acro{RS}{reliable subcarriers}
\acro{URS}{unreliable subcarriers}
\acro{AE-DNN}{auto-encoder deep neural network}
\acro{AE}{auto-encoder}
\acro{T-DFT}{truncated discrete Fourier transform}
\acro{TA-TDFT}{temporal averaging T-DFT}
\acro{TA}{time averaging}
\acro{PDP}{power delay profile}
\acro{1G}{first generation}
\acro{2G}{second generation}
\acro{3G}{third generation}
\acro{3GPP}{Third Generation Partnership Project}
\acro{4G}{fourth generation}
\acro{5G}{fifth generation}
\acro{802.11}{IEEE 802.11 specifications}
\acro{A/D}{analog-to-digital}
\acro{ADC}{analog-to-digital}
\acro{AM}{amplitude modulation}
\acro{AP}{access point}
\acro{AR}{augmented reality}
\acro{ASIC}{application-specific integrated circuit}
\acro{ASIP}{Application Specific Integrated Processors}
\acro{AWGN}{additive white Gaussian noise}
\acro{BCJR}{Bahl, Cocke, Jelinek and Raviv}
\acro{BER}{bit error rate}
\acro{BFDM}{bi-orthogonal frequency division multiplexing}
\acro{BPSK}{binary phase shift keying}
\acro{BS}{base stations}
\acro{CA}{carrier aggregation}
\acro{CAF}{cyclic autocorrelation function}
\acro{Car-2-x}{car-to-car and car-to-infrastructure communication}
\acro{CAZAC}{constant amplitude zero autocorrelation waveform}
\acro{CB-FMT}{cyclic block filtered multitone}
\acro{CCDF}{complementary cumulative density function}
\acro{CDF}{cumulative density function}
\acro{CDMA}{code-division multiple access}
\acro{CFO}{carrier frequency offset}
\acro{CIR}{channel impulse response}
\acro{CM}{complex multiplication}
\acro{COFDM}{coded-\acs{OFDM}}
\acro{CoMP}{coordinated multi point}
\acro{COQAM}{cyclic OQAM}
\acro{CP}{cyclic prefix}
\acro{CR}{cognitive radio}
\acro{CRC}{cyclic redundancy check}
\acro{CRLB}{Cram\'{e}r-Rao lower bound}
\acro{CS}{cyclic suffix}
\acro{CSI}{channel state information}
\acro{CSMA}{carrier-sense multiple access}
\acro{CWCU}{component-wise conditionally unbiased}
\acro{D/A}{digital-to-analog}
\acro{D2D}{device-to-device}
\acro{DAC}{digital-to-analog}
\acro{DC}{direct current}
\acro{DFE}{decision feedback equalizer}
\acro{DFT}{discrete Fourier transform}
\acro{DL}{deep learning}
\acro{DMT}{discrete multitone}
\acro{DNN}{deep neural network}
\acro{FNN}{feed-forward neural network}
\acro{DSA}{dynamic spectrum access}
\acro{DSL}{digital subscriber line}
\acro{DSP}{digital signal processor}
\acro{DTFT}{discrete-time Fourier transform}
\acro{DVB}{digital video broadcasting}
\acro{DVB-T}{terrestrial digital video broadcasting}
\acro{DWMT}{discrete wavelet multi tone}
\acro{DZT}{discrete Zak transform}
\acro{E2E}{end-to-end}
\acro{eNodeB}{evolved node b base station}
\acro{E-SNR}{effective signal-to-noise ratio}
\acro{EVD}{eigenvalue decomposition}
\acro{FBMC}{filter bank multicarrier}
\acro{FD}{frequency-domain}
\acro{FDD}{frequency-division duplexing}
\acro{FDE}{frequency domain equalization}
\acro{FDM}{frequency division multiplex}
\acro{FDMA}{frequency-division multiple access}
\acro{FEC}{forward error correction}
\acro{FER}{frame error rate}
\acro{FFT}{fast Fourier transform}
\acro{FIR}{finite impulse response}
\acro{FM}		{frequency modulation}
\acro{FMT}{filtered multi tone}
\acro{FO}{frequency offset}
\acro{F-OFDM}{filtered-\acs{OFDM}}
\acro{FPGA}{field programmable gate array}
\acro{FSC}{frequency selective channel}
\acro{FS-OQAM-GFDM}{frequency-shift OQAM-GFDM}
\acro{FT}{Fourier transform}
\acro{FTD}{fractional time delay}
\acro{FTN}{faster-than-Nyquist signaling}
\acro{GFDM}{generalized frequency division multiplexing}
\acro{GFDMA}{generalized frequency division multiple access}
\acro{GMC-CDM}{generalized	multicarrier code-division multiplexing}
\acro{GNSS}{global navigation satellite system}
\acro{GS}{guard symbols}
\acro{GSM}{Groupe Sp\'{e}cial Mobile}
\acro{GUI}{graphical user interface}
\acro{H2H}{human-to-human}
\acro{H2M}{human-to-machine}
\acro{HTC}{human type communication}
\acro{I}{in-phase}
\acro{i.i.d.}{independent and identically distributed}
\acro{IB}{in-band}
\acro{IBI}{inter-block interference}
\acro{IC}{interference cancellation}
\acro{ICI}{inter-carrier interference}
\acro{ICT}{information and communication technologies}
\acro{ICV}{information coefficient vector}
\acro{IDFT}{inverse discrete Fourier transform}
\acro{IDMA}{interleave division multiple access}
\acro{IEEE}{institute of electrical and electronics engineers}
\acro{IF}{intermediate frequency}
\acro{IFFT}{inverse fast Fourier transform}
\acro{IoT}{Internet of Things}
\acro{IOTA}{isotropic orthogonal transform algorithm}
\acro{IP}{internet protocole}
\acro{IP-core}{intellectual property core}
\acro{ISDB-T}{terrestrial integrated services digital broadcasting}
\acro{ISDN}{integrated services digital network}
\acro{ISI}{inter-symbol interference}
\acro{ITU}{International Telecommunication Union}
\acro{IUI}{inter-user interference}
\acro{LAN}{local area netwrok}
\acro{LLR}{log-likelihood ratio}
\acro{LMMSE}{linear minimum mean square error}
\acro{LNA}{low noise amplifier}
\acro{LO}{local oscillator}
\acro{LOS}{line-of-sight}
\acro{LP}{low-pass}
\acro{LPF}{low-pass filter}
\acro{LS}{least squares}
\acro{LTE}{Long Term Evolution}
\acro{LTE-A}{LTE-Advanced}
\acro{LTIV}{linear time invariant}
\acro{LTV}{linear time-variant}
\acro{LUT}{lookup table}
\acro{M2M}{machine-to-machine}
\acro{MA}{multiple access}
\acro{MAC}{multiple access control}
\acro{MAP}{maximum a posteriori}
\acro{MC}{multicarrier}
\acro{MCA}{multicarrier access}
\acro{MCM}{multicarrier modulation}
\acro{MCS}{modulation coding scheme}
\acro{MF}{matched filter}
\acro{MF-SIC}{matched filter with successive interference cancellation}
\acro{MIMO}{multiple-input, multiple-output}
\acro{MISO}{multiple-input single-output}
\acro{ML}{machien learning}
\acro{MLD}{maximum likelihood detection}
\acro{MLE}{maximum likelihood estimator}
\acro{MMSE}{minimum mean squared error}
\acro{MRC}{maximum ratio combining}
\acro{MS}{mobile stations}
\acro{MSE}{mean squared error}
\acro{MSK}{Minimum-shift keying}
\acro{MSSS}[MSSS]	{mean-square signal separation}
\acro{MTC}{machine type communication}
\acro{MU}{multi user}
\acro{MVUE}{minimum variance unbiased estimator}
\acro{NEF}{noise enhancement factor}
\acro{NLOS}{non-line-of-sight}
\acro{NMSE}{normalized mean-squared error}
\acro{NOMA}{non-orthogonal multiple access}
\acro{NPR}{near-perfect reconstruction}
\acro{NRZ}{non-return-to-zero}
\acro{OFDM}{orthogonal frequency division multiplexing}
\acro{OFDMA}{orthogonal frequency division multiple access}
\acro{OOB}{out-of-band}
\acro{OQAM}{offset quadrature amplitude modulation}
\acro{OQPSK}{offset quadrature phase shift keying}
\acro{OTFS}{orthogonal time frequency space}
\acro{PA}{power amplifier}
\acro{PAM}{pulse amplitude modulation}
\acro{PAPR}{peak-to-average power ratio}
\acro{PC-CC}{parallel concatenated convolutional code}
\acro{PCP}{pseudo-circular pre/post-amble}
\acro{PD}{probability of detection}
\acro{pdf}{probability density function}
\acro{PDF}{probability distribution function}

\acro{PFA}{probability of false alarm}
\acro{PHY}{physical layer}
\acro{PIC}{parallel interference cancellation}
\acro{PLC}{power line communication}
\acro{PMF}{probability mass function}
\acro{PN}{pseudo noise}
\acro{ppm}{parts per million}
\acro{PRB}{physical resource block}
\acro{PRB}{physical resource block}
\acro{PSD}{power spectral density}
\acro{Q}{quadrature-phase}
\acro{QAM}{quadrature amplitude modulation}
\acro{QoS}{quality of service}
\acro{QPSK}{quadrature phase shift keying}
\acro{R/W}{read-or-write}
\acro{RAM}{random-access memmory}
\acro{RAN}{radio access network}
\acro{RAT}{radio access technologies}
\acro{RC}{raised cosine}
\acro{RF}{radio frequency}
\acro{rms}{root mean square}
\acro{RRC}{root raised cosine}
\acro{RW}{read-and-write}
\acro{SC}{single-carrier}
\acro{SCA}{single-carrier access}
\acro{SC-FDE}{single-carrier with frequency domain equalization}
\acro{SC-FDM}{single-carrier frequency division multiplexing}
\acro{SC-FDMA}{single-carrier frequency division multiple access}
\acro{SD}{sphere decoding}
\acro{SDD}{space-division duplexing}
\acro{SDMA}{space division multiple access}
\acro{SDR}{software-defined radio}
\acro{SDW}{software-defined waveform}
\acro{SEFDM}{spectrally efficient frequency division multiplexing}
\acro{SE-FDM}{spectrally efficient frequency division multiplexing}
\acro{SER}{symbol error rate}
\acro{SIC}{successive interference cancellation}
\acro{SINR}{signal-to-interference-plus-noise ratio}
\acro{SIR}{signal-to-interference ratio}
\acro{SISO}{single-input, single-output}
\acro{SMS}{Short Message Service}
\acro{SNR}{signal-to-noise ratio}
\acro{STC}{space-time coding}
\acro{STFT}{short-time Fourier transform}
\acro{STO}{symbol time offset}
\acro{SU}{single user}
\acro{SVD}{singular value decomposition}
\acro{TD}{time-domain}	
\acro{TDD}{time-division duplexing}
\acro{TDMA}{time-division multiple access}
\acro{TFL}{time-frequency localization}
\acro{TO}{time offset}
\acro{TS-OQAM-GFDM}{time-shifted OQAM-GFDM}
\acro{UE}{user equipment}
\acro{UFMC}{universally filtered multicarrier}
\acro{UL}{uplink}
\acro{US}{uncorrelated scattering}
\acro{USB}{universal serial bus}
\acro{UW}{unique word}
\acro{VLC}{visible light communications}
\acro{VR}{virtual reality}
\acro{WCP}{windowing and \acs{CP}}	
\acro{WHT}{Walsh-Hadamard transform}
\acro{WiMAX}{worldwide interoperability for microwave access}
\acro{WLAN}{wireless local area network}
\acro{W-OFDM}{windowed-\acs{OFDM}}	
\acro{WOLA}{windowing and overlapping}	
\acro{WSS}{wide-sense stationary}
\acro{ZCT}{Zadoff-Chu transform}
\acro{ZF}{zero-forcing}
\acro{ZMCSCG}{zero-mean circularly-symmetric complex Gaussian}
\acro{ZP}{zero-padding}
\acro{ZT}{zero-tail}
\end{acronym}

\usepackage[utf8]{inputenc}
\usepackage[english]{babel}
\usepackage{amsthm}
\usepackage{commath}

\begin{document}


\title{A Survey on Deep Learning based Channel Estimation in Doubly Dispersive Environments}
\author{Abdul Karim Gizzini, \IEEEmembership{Member, IEEE},
Marwa Chafii, \IEEEmembership{Member, IEEE}

\thanks{

Authors acknowledge the CY INEX for the support of the project through the ASIA Chair of Excellence Grant (PIA/ANR-16-IDEX-0008).

Abdul Karim Gizzini is with ETIS, UMR8051, CY Cergy Paris Université, ENSEA, CNRS, France (e-mail: abdulkarim.gizzini@ensea.fr).

 Marwa Chafii is with the Engineering Division, New York University (NYU) Abu Dhabi, 129188, UAE, and NYU WIRELESS, NYU Tandon School of Engineering, Brooklyn, 11201, NY (e-mail: marwa.chafii@nyu.edu).

}

}
%
\markboth{}
{}
%
%
\maketitle
\begin{abstract}
 Wireless communications systems are impacted by multi-path fading and Doppler shift in dynamic environments, where the channel becomes doubly-dispersive and its estimation becomes an arduous task. Only a few pilots are used for channel estimation in conventional approaches to preserve high data rate transmission. Consequently, such estimators experience a significant performance degradation in high mobility scenarios. Recently, deep learning has been employed for  doubly-dispersive channel estimation due to its low-complexity, robustness, and good generalization ability. Against this backdrop, the current paper presents a comprehensive survey on channel estimation techniques based on deep learning by deeply investigating  different methods. The study also provides extensive experimental simulations followed by a computational complexity analysis. After considering different parameters such as modulation order, mobility, frame length, and deep learning architecture, the performance of the studied estimators is evaluated in several mobility scenarios. In addition, the source codes are made available online in order to make the results reproducible.
\end{abstract}

\section{Introduction} \label{introduction}
\IEEEPARstart{W}ith the commercialization of fifth generation networks globally, research into sixth generation (6G) networks has been initiated to address the demands for high data rates and low latency mobile applications, including unmanned aerial vehicles~\cite{ref_UAV}, high-speed railway~\cite{ref_RW}, and vehicular communications~\cite{ref_VCC}.
Mobile wireless communications systems offer the freedom to move around without being disconnected from the network. However, the mobility feature is ridden with several challenges that have a severely adverse impact on the communication reliability, such as fast and frequent handovers~\cite{ref_FHO}, carrier frequency offset~\cite{ref_FOE}, inter-carrier interference~\cite{ref_ICInt}, high penetration loss~\cite{ref_HPL}, and fast time-varying wireless channel~\cite{bomfin2021robust}. 

In wireless environment, transmitted signals are known to propagate via a multitude of paths, each entailing a different attenuation and delay in addition to the Doppler shift effect stemming from the motion of network nodes along with the surrounding environment. As a result, the wireless channel becomes frequency-selective and time-varying. Given that a precisely estimated channel response influences the follow-up equalization, demodulation, and decoding operations at the receiver, the accuracy of the channel estimation influences the system performance. Therefore, ensuring communication reliability via accurate channel estimation in such environments is highly important.

In the extant literature, a vast body of work has been carried out to address the problem of doubly-dispersive channels. While some works have focused on investigating the waveform design~\cite{ref_Roberto_Receiver,ref_roberto_additional1, ref_nimr_additional2,ref_marwa_additional3}, we are interested in this paper in the channel estimation task. In general, channel estimators can be classified into two main categories: (\textit{i}) {\ac{SBS}} channel estimators: the channel is estimated for each received symbol separately using only the previous and current received pilots~\cite{ref_sta,ref_cdp,ref_trfii} (\textit{ii}) \ac{FBF} channel estimators: the previous, existing, as well as future pilots are employed in the channel estimation for each received symbol~\cite{ehsanfar2020uw}. It is possible to achieve a higher channel estimation accuracy by utilizing {\ac{FBF}} estimators, since the channel estimation of each symbol benefits from the combined knowledge of all allocated pilots within the frame. However, the conventional estimators' performance mainly relies on the  allocated reference training pilots within the transmitted frames. The majority of standards allocate a few pilots to maintain a good transmission data rate. Therefore, these pilots are insufficient for accurately tracking the doubly-dispersive channel, because they are not spaced closely enough to capture the variation of the channel in the frequency domain. Consequently, conventional estimators are primarily based on the demapped data subcarriers, besides pilot subcarriers to update the channel estimate for each received symbol. This procedure called {\ac{DPA}} channel estimation is regarded as unreliable because the demapping error gets enlarged from one symbol to another, which leads to another additional error in the estimation process, especially in highly dynamic time-varying channels. Moreover, other  conventional estimators like the \ac{LMMSE}~\cite{ref_lmmse} estimator
rely on many assumptions that limit their performance in highly dynamic time-varying
channels. Moreover, linear conventional estimators are impractical solutions in real case scenarios as they rely on statistical models and require high implementation complexity, in addition, they lack robustness in highly dynamic environments. Therefore, investigating estimators with a
good trade-off complexity vs. performance is a crucial need for improving the channel estimation accuracy while preserving good data rate as well as maintaining  affordable computational complexity.


As a prevailing approach to AI, \ac{DL} is an efficient method to analyze data by identifying patterns and learning underlying structures, denoting an effective approach to problems faced in various scientific fields. {\ac{DL}} algorithms have been integrated into the physical layer of wireless communications systems  ~\cite{ref_DL_PHY1, ref_DL_PHY2,chafii2018enhancing}, including channel estimation~\cite{ref_DL_Chest1,ref_DL_Chest2,ref_DL_Chest3,additional_ref1,additional_ref2,additional_ref3}. In turn, this is attributable to the great success in enhancing the overall system performance, particularly when used in addition to conventional estimators, where coarse channel estimation is derived from conventional estimators, following which DL is employed to achieve a fine estimation. Therefore, DL-based channel estimators are capable of significantly enhancing the performance while preserving low computational complexity. 
In addition, the GPU-based distributed processing allows the DL employment in real-time applications, as a result of which DL can overcome the limitations of traditional channel estimation through robust, low-complexity, and generalized solutions that improve the performance of wireless systems.

Motivated by these advantages, {\ac{DL}} algorithms have been integrated in frequency-selective~\cite{additional_ref1,additional_ref2,additional_ref3} and doubly-dispersive channel estimation. In this survey, we examine the recently proposed DL-based channel estimation schemes in doubly-dispersive environments, where DL algorithms are utilized in two different manners: (\textit{i})  \acp{FNN} with different architectures and configurations are employed on top of the conventional {\ac{SBS}} channel estimators~\cite{ref_dpa_dnn,ref_STA_DNN,ref_TRFI_DNN}. (\text{ii}) {\acp{CNN}} processing is employed where the estimated channel for the entire frame is modeled as a 2D low-resolution noisy image, whereas \ac{CNN}-based processing is implemented as super resolution and denoising techniques~\cite{ref_ChannelNet,ref_TS_ChannelNet,ref_WI_CNN}. 

The majority of surveys conducted in the literature ~\cite{ref_mimo2,refsur} lack intensive simulations in the performance evaluation and complexity analysis of the studied channel estimators. Moreover, they do not cover both SBS and FBF based estimators. In addition,~\cite{ref_mimo2} compares the performance of different DL architectures used after the {\ac{LS}} and the LMMSE estimators without considering several conventional channel estimation schemes, whereas~\cite{refsur} provides a general overview of several channel estimators without any performance evaluation. Given this context, to the best of our knowledge, this is the first survey that presents a comprehensive study on the recently proposed DL-based SBS and FBF estimators in doubly-dispersive environments, while presenting intensive simulations evaluating the system performance in different scenarios, providing a detailed complexity analysis, as well as the source codes to reproduce all the presented results.
We believe that this survey is a very relevant reference to initiate researches pertaining to the domain of deep learning based channel estimation in doubly dispersive channels.
The contributions of this paper can be summarized in the following manner  
\begin{itemize}
    \item Comprehensive study on the recently proposed {\ac{DL}}-based channel estimation techniques for doubly-dispersive channels.
    
    \item Overview of the {\ac{DL}} networks, especially those used in the studied channel estimators, such as \ac{FNN}, {\ac{LSTM}}, {\ac{SR-CNN}}, and {\ac{DN-CNN}}.
    
    \item Performance analysis of different channel estimation schemes and a fair comparison between them in terms of {\ac{NMSE}} and {\ac{BER}} for different mobility scenarios and frame length, and modulation order.
    
    \item Detailed computational complexity analysis for the studied channel estimators concerning the overall required real-valued operations.
    
    \item Simulation source code for various channel estimation schemes to reproduce all the comparison results presented in this paper~\cite{ref_githubcodes}.
    
\end{itemize}


The remainder of this paper is organized as follows: Section~\ref{system_model} elucidates the system model, illustrating signal transmission over a doubly-dispersive channel. Section~\ref{DL_overview} provides a brief overview of the main {\ac{DL}} networks employed in this survey. The recently proposed {\ac{DL}}-based {\ac{SBS}} and {\ac{FBF}} channel estimation schemes are thoroughly investigated and discussed in Sections~\ref{sbs_estimators} and~\ref{fbf_estimators}, respectively. In Section~\ref{simulation_results}, different modulation orders are used to present simulation results, wherein the performance of the studied estimators is examined in terms of \ac{BER}, and \ac{NMSE}. Detailed computational complexity analysis is provided in Section~\ref{complexity}. Finally, Section~\ref{conclusions} concludes this study.

\textbf{Notations}: Throughout the paper, vectors are defined
with lowercase bold symbols $\ma{x}$ whose $k$-th element is $\ma{x}[k]$. Time and frequency domain vectors are represented by $\ma{x}$ and $\tilde{\ma{x}}$ respectively. Matrices are written as uppercase bold symbols $\ma{X}$. $\Ex{.}$ denotes the expectation operator. The trace of a square matrix $\ma{X}$ is $\trace{\ma{X}}$. The notation $\odot$ and $\oslash$ refer to the element-wise multiplication and division operations, respectively. Finally. the pseudo inverse and conjugate matrices of $\ma{X}$ are signified by $\ma{X}^{\dagger}$ and $\ma{X}^{\text{H}}$, respectively.

\section{System Model} \label{system_model}

Consider a frame comprising $I$ {\ac{OFDM}} symbols.  The $i$-th transmitted frequency-domain {\ac{OFDM}} symbol  $\tilde{\ma{x}}_i[k]$, is denoted by
\begin{equation}
   \tilde{\ma{x}}_i[k] = \left\{
            \begin{array}{ll}        
                \tilde{\ma{x}}_{\text{d}_{i}}[k],&\quad k \in \Kd. \\
                \tilde{\ma{x}}_{\text{p}_{i}}[k],&\quad k \in \Kp. \\
            \end{array}\right.
\label{eq: xK}
\end{equation}
where $0 \leq k \leq K - 1$. $ \tilde{\ma{x}}_{\text{d}_{i}}[k]$ and $ \tilde{\ma{x}}_{\text{p}_{i}}[k]$ represent the modulated data symbols and the predefined pilot symbols allocated at a set of subcarriers denoted $\Kd$ and $\Kp$, respectively. $\ma{x}_i[k]$ is converted to the time domain by applying the \ac{IDFT}, such that
\begin{equation}
\ma{x}_i[n] = \frac{1}{\sqrt{{{K}}}} \sum_{k=0}^{{K}-1} \tilde{\ma{x}}_i[k] e^{j2\pi\frac{ nk}{{K}}}.  
\label{eq:x_time}
\end{equation}%
A \ac{CP} of length larger than the delay spread is added. Therefore, after passing via the doubly-dispersive channel and removing the \ac{CP}, the received {\ac{OFDM}} symbol $\ma{y}_i[n]$ can be expressed as follows 
\begin{equation}
\begin{split}
\ma{y}_i[n] &=  \sum_{l=0}^{L-1} {\ma{h}_i[l,n]} \ma{x}_i[n-l] + {\ma{v}}_i[n]\\  
             &= \frac{1}{\sqrt{{K}}} \sum_{k=0}^{{K}-1}  {\tilde{\ma{h}}_i[k,n]} \tilde{\ma{x}}_i[k] e^{j2\pi\frac{ nk}{{K}}} + {\ma{v}}_i[n]. 
\end{split}
\label{eq:y_time_channel}
\end{equation}

$\ma{h}_i[l,n]$ denotes the delay-time response of the discrete \ac{LTV} channel of $L$ taps at the $i$-th \ac{OFDM} symbol, whereas   $\tilde{\ma{h}}_i[k,n] = \sum_{l=0}^{L-1} {{\ma{h}}_i[l,n]} e^{-j2\pi\frac{ lk}{{K}}} $ refers to the frequency-time response. Moreover, ${\ma{v}}_i$ signifies the \ac{AWGN} of variance $\sigma^2$.
The $i$-th received frequency-domain {\ac{OFDM}} symbol is derived from \eqref{eq:y_time_channel} via {\ac{DFT}}, and thus
\begin{equation}
\begin{split}
\tilde{\ma{y}}_i[k] &=
                     \frac{1}{{{K}}} \sum_{q=0}^{{K}-1} \tilde{\ma{x}}_i[q] \sum_{n=0}^{{K}-1} \tilde{\ma{h}}_i[q,n] e^{-j2\pi\frac{ n (k-q)}{{K}}}+ \tilde{\ma{v}}_i[k].
\end{split}\label{eq:y_frquency}
\end{equation}

It is noteworthy that index $k$ is used in~{\eqref{eq:y_time_channel}} to express the channel delay-time response in terms of the channel frequency-time response. While the change of index into $q$ in~{\eqref{eq:y_frquency}} is used to express the $i$-th received symbol in frequency domain. This, in turn, better illustrates the DFT transform. Moreover, $\tilde{\ma{h}}_i[q,n]$  refers to time-variant at the scale of the \ac{OFDM} symbol duration (the index $i$) and within the symbol itself (the index $n$). Accordingly, 
\begin{equation}
	\begin{split}
		\tilde{\ma{h}}_i[q,n] &= \sum_{l=0}^{L-1} e^{-j2\pi \frac{lq}{K}}\int_{\nu = -\nu_d}^{\nu = \nu_d} \bar{h}(l, \nu) e^{j2\pi \nu n_i} e^{j2\pi \nu n}d\nu, \label{eq: delay-doppler}
	\end{split}
\end{equation}
where $\bar{h}(l, \nu) = \sum\limits_{n} {h}[l,n] e^{-j2\pi n\nu} $ signifies the channel delay-Doppler response, $\nu$ refers to the  normalized Doppler frequency,  $n_i = i(K+K_{\text{cp}}) + K_{\text{cp}}$. And $\nu_d = \frac{f_d}{F_s}$ represents the maximum  Doppler frequency. Let
\begin{equation}
	\begin{split}
\bar{\ma{h}}_i[l,v]	&=\frac{1}{K}\sum_{q=0}^{{K}-1}\sum_{n=0}^{{K}-1} \tilde{\ma{h}}_i[q,n] e^{-j2\pi\frac{ n v }{K}}e^{j2\pi\frac{ q l }{K}}\\
&=  \int_{\nu = -\nu_d}^{\nu = \nu_d} \bar{h}(l, \nu) e^{j2\pi \nu n_i}\sum_{n=0}^{{K}-1} e^{-j2\pi (\nu-\frac{v}{K})
	 n}d\nu,
	\end{split}
	\label{eq:DD}
\end{equation}
be the discrete delay-Doppler response at the $i$-th {\ac{OFDM}} symbol. For the sake of simplicity, $\bar{h}(l,\nu)$ is assumed to be uncorrelated in both domains~\cite{MATZ20111}, such that $\Ex{\bar{h}(l,\nu)\bar{h}^*(l',\nu')} = S_h(l,\nu)\delta(l-l')\delta(\nu-\nu')$, where  $S_h(l,\nu)$ is the delay-Doppler spectrum~\cite{ref_WC}, and $\delta(x)$ denotes the Dirac delta function. Using~\eqref{eq:DD}, we have
\begin{equation}
	\begin{split}
	\Ex{\bar{\ma{h}}_i[l,v]\bar{\ma{h}}_i^*[l,v']} = &  \\	
		  \int_{\nu = -\nu_d}^{\nu = \nu_d} S_h(l,\nu)\sum_{n=0}^{{K}-1}&\sum_{n'=0}^{{K}-1} e^{-j2\pi \nu (n-n')} 
			  e^{-j2\pi \frac{n'v'-nv}{K}}
		d\nu .
	\end{split}
\end{equation}
This correlation that is independent of the index $i$ can be approximated as follows
\begin{equation}
	\begin{split}
		\Ex{\bar{\ma{h}}_i[l,v]\bar{\ma{h}}_i^*[l,v']} &\approx K^2\rho[l,v]\delta[v-v'],\\
		\mbox{where } & \rho[l,v] = S_h(l, \frac{v}{N}). \label{eq:correlation}
	\end{split}
\end{equation} 
The time selectivity of the channel depends on the mobility. In very low mobility, where  $f_{\text{d}} \approx 0$, $\tilde{\ma{h}}_i[q,n] = \tilde{\ma{h}}[q]$ is constant  during the whole frame. For moderate to high mobility, the channel variation within the duration of one \ac{OFDM} symbol is negligible, and  therefore,  $\tilde{\ma{h}}_i[q,n] = \tilde{\ma{h}}_i[q]$. At very high mobility, the channel becomes variant within a single \ac{OFDM} symbol. In this instance, $\tilde{\ma{h}}_i[q,n] = \tilde{\ma{h}}_i[q] + \tilde{\ma{\epsilon}}_i[q,n] $, where
\begin{equation}
 \tilde{\ma{h}}_i[q] = \frac{1}{K}\sum_{n=0}^{K-1}\tilde{\ma{h}}_i[q,n],~ \mbox{and } \tilde{\ma{\epsilon}}_i[q,n] = \tilde{\ma{h}}_i[q,n]- \tilde{\ma{h}}_i[q]. \label{eq:channel_for_estimation}
\end{equation}
Replacing this in \eqref{eq:y_frquency}, we get
\begin{equation}
	\begin{split}
		\tilde{\ma{y}}_{{i}}[k] 
		&= \tilde{\ma{h}}_i[k] \tilde{\ma{x}}_i[k] + \tilde{\ma{e}}_{i,\text{d}}[k] + \tilde{\ma{v}}_i[k],~ k \in \Kon.
	\end{split}            
	\label{eq: system_model}
\end{equation}
The term $\tilde{\ma{e}}_{i,\text{d}}[k]$ denotes the Doppler interference given by 
\begin{equation}
	\begin{split}
		\tilde{\ma{e}}_{i,\text{d}}[k] &=  \frac{1}{{{K}}} \sum_{\substack{q=0 \\ q \neq k}}^{{K}-1} \sum_{n=0}^{{K}-1} \tilde{\ma{h}}_i[q,n] e^{-j2\pi\frac{ n (k-q)}{{K}}}\tilde{\ma{x}}_i[q]\\
		&= \frac{1}{K}\sum_{\substack{q\in \Kon \\ q \neq k}} \sum_{l=0}^{L-1}\bar{\ma{h}}_i[l,k-q] e^{-j2\pi \frac{lq}{K}} \tilde{\ma{x}}_i[q].
		\label{eq: Doppler}
	\end{split}
\end{equation}
%
%
The Doppler interference destroys the orthogonality of the subcarriers within the received {\ac{OFDM}} symbol, leading to a significant degradation in the overall system performance~\cite{ref19}.
Assuming that the subcarriers are uncorrelated with power $E_q$, i.e. $\Ex{\tilde{\ma{x}}_i[q]\tilde{\ma{x}}^*_i[q']} = E_q \delta[q-q']$ and using \eqref{eq:correlation} then 
\begin{equation}
	\begin{split}
		\Ex{\tilde{\ma{e}}_{i,\text{d}}[k]\tilde{\ma{e}}^*_{i,\text{d}}[k']} &= \sum_{l=0}^{L-1}\sum_{\substack{q\in \Kon \\ q \neq k}}E_q \rho[l,k-q]\delta[k-k']\\
		& = \sigma^2_d[k] \delta[k-k'].
	\end{split} \label{eq:Doppler correlation}
\end{equation}
Thus, it is assumed that the Doppler interference is uncorrelated. However, the variance  $\sigma^2_d[k] = \Ex{|\tilde{\ma{e}}_{i,\text{d}}[k]|^2}$ depends on the subcarrier index. Noting that 
\begin{equation}
	\tilde{\ma{h}}_i[k] = \frac{1}{K}\sum_{l=0}^{L-1}\bar{\ma{h}}_i[l,0]e^{-j2\pi \frac{kl}{K}}, \label{eq:channel_for_estimation_with L}
\end{equation}
the channel gain and Doppler interference are uncorrelated, i.e. $\Ex{\tilde{\ma{h}}_i[k]  \tilde{\ma{e}}_{i,\text{d}}^*[k] } = 0$. Moreover, it is possible to estimate the $\tilde{\ma{h}}_i[k]$ from $L$ uncorrelated taps defined by $\bar{\ma{h}}_i[l,0]$.

\section{DL Techniques Overview} \label{DL_overview}

This section discusses the {\ac{DL}} networks employed in the studied {\ac{DL}}-based channel estimation schemes, providing the mathematical representation of each network.

\subsection{FNN}

Neural networks are one of the most popular machine learning algorithms~\cite{ref_dnn}. Initially, neural networks are inspired by the neural architecture of a human brain, For this reason, the basic building block is called a neuron as is the case with a human brain. Its functionality is similar to that of a human neuron, i.e. it takes in some inputs and then fires an output. In purely mathematical terms, a neuron denotes a placeholder for a mathematical function whose job is to yield an output by applying the function on the given inputs. Neurons are stacked together to form a layer. The neural network comprises at least one layer;  in case multiple layers are employed, the neural network is referred to as deep {\ac{FNN}}.

\begin{figure}[t]
    \centering
    \includegraphics[width=1\columnwidth]{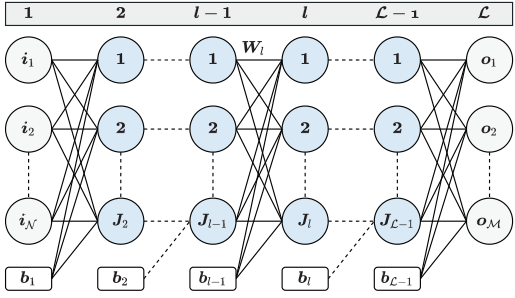}
    \caption{FNN architecture showing the input, output, and hidden layers.}
    \label{fig:DNN}
\end{figure}

Consider a {\ac{FNN}} architecture shown in \figref{fig:DNN}. Here $\mathcal{L}$ represents the number of layers, including one input layer, $\mathcal{L}-2$ hidden layers, as well as one output layer . The $l$-th hidden layer of the network consists of $J_{l}$ neurons where $2\leq l \leq L-1$. Moreover, each neuron in the $l$-th hidden layer is denoted by $j$ where $j$ $1\leq j \leq J_{l}$. The {\ac{FNN}} inputs $\ma{i}$ and outputs $\ma{o}$ are expressed as $\ma{i} = [i_{1}, i_{2},...,i_{\mathcal{N}}]^T \in \mathbb{R}^{\mathcal{N} \times 1}$ and $\ma{o} = [o_{1}, o_{2},..., o_{\mathcal{M}}]^T \in \mathbb{R}^{\mathcal{M} \times 1}$, where $\mathcal{N}$ and $\mathcal{M}$ refer to the number of {\ac{FNN}} inputs and outputs, respectively.
$\ma{W}_{l} \in \mathbb{R}^{J_{l} \times J_{l-1}}$, and $\ma{b}_{l} \in \mathbb{R}^{J_{l} \times 1}$ are used to express the weight matrix and the bias vector of the $l$-th hidden layer, respectively.

Each neuron $n_{(l,j)}$ performs a nonlinear transform of a weighted summation of the preceding layer's output values. This nonlinear transformation is represented by the activation function ${f}_{(l,j)}$ on the neuron input vector $\ma{i}_{(l)} \in \mathbb{R}^{J_{l-1} \times 1}$ using its weight vector $\ma{\omega}_{(l,j)} \in \mathbb{R}^{J_{l-1} \times 1} $, and bias ${b}_{(l,j)}$, respectively. The neuron's output ${o}_{(l,j)}$ is
\begin{equation}
{o}_{(l,j)} = {f}_{(l,j)} \Big( b_{(l,j)} + {\ma{\omega}^T_{(l,j)}}  \ma{i}_{(l)} \Big).
\end{equation}
The \ac{DNN} overall output of the $l$-th hidden layer is signified by the vector form
\begin{equation}
\ma{o}_{(l)} = \ma{f}_{(l)} \Big( \ma{b}_{(l)} + {\ma{W}_{(l)}} \ma{i}_{(l)} \Big),~\ma{i}_{(l+1)} = \ma{o}_{(l)},
\label{eq:hiddenlayeroutput}
\end{equation}
where $\ma{f}_{(l)}$ is a vector resulting from the stacking of the $n_{l}$ activation functions.

After the selection of the \ac{FNN} architecture, the parameter ${\theta} = (\ma{W},\ma{B})$ representing the total \ac{FNN} weights and biases must be estimated via the learning procedure applied during the \ac{FNN} training phase. As well known, ${\theta}$ estimation is obtained by minimizing a loss function $\text{Loss}({\theta})$. The loss function measures how far apart the predicted \ac{FNN} outputs ($\ma{o}_{(\mathcal{L})}^{\text{(P)}}$) are from the true outputs ($\ma{o}_{(\mathcal{L})}^{\text{(T)}}$). Therefore, the \ac{FNN} training phase carried over $N_{\text{train}}$ training samples can be explained in two steps: (i) calculate the loss, and (ii) update ${\theta}$. This process is  repeated until convergence, so that the loss becomes very small. Accordingly, various optimization algorithms can be used for minimizing $\text{Loss}({\theta})$ by iteratively updating the parameter ${\theta}$, i.e., stochastic gradient descent~\cite{ref_dnn}, root mean square prop~\cite{ref_dnn_rmsp}, and adaptive moment estimation (ADAM)~\cite{ref_dnn_optimizers}.

%
%
%
%
%
%

%
%
The final step after \ac{FNN} training is to test the trained \ac{FNN} on new data in order to evaluate its performance. An elaborate comprehensive analysis of \ac{FNN} different principles is presented in~\cite{ref_dnn_comprehansive}.
\subsection{LSTM}
Another well-known {\ac{DL}} tool is available in the form of {\ac{LSTM}} networks that essentially deal with sequential data where the order of the data matters and a correlation exists between the previous and the future data. In this context, {\ac{LSTM}} networks are defined with a special architecture capable of learning the data correlation over time, which enables the {\ac{LSTM}} network to predict the future data based on prior observations.

\begin{figure}[t]
	\centering
  \includegraphics[width=\columnwidth]{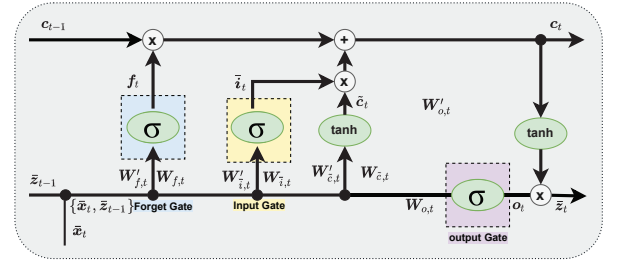}
  \caption{LSTM unit architecture~\cite{gizzini:hal-03365697}.}
  \label{fig:lstm_archi}
\end{figure}

The {\ac{LSTM}} unit, as shown in \figref{fig:lstm_archi}, contains computational blocks referred to as gates, which are responsible for controlling and tracking the information flow over time. The {\ac{LSTM}} network mechanism can be explicated in four major steps:
  \paragraph{Forget the irrelevant information} In general, the {\ac{LSTM}} unit classifies the input data into relevant and irrelevant information. The first processing step entails eliminating the irrelevant information that is not important for predicting the future. This can be undertaken through the forget gate that decides which information the {\ac{LSTM}} unit should retain, and which information can be deleted. The forget gate processing is defined as follows 
  
    \begin{equation}
    {\ma{f}}_{t} = \sigma (\ma{W}_{f, t}\bar{\ma{x}}_{t} + \ma{W}^{\prime}_{f,t}\bar{\ma{z}}_{t-1} + \bar{\ma{b}}_{f,t}),
    \label{eq: lstm_fg}
    \end{equation}
    where $\bar{\sigma}$ denotes the sigmoid function, $\ma{W}_{f,t} \in \mathbb{R}^{P \times K_{in}}$,  $\ma{W}^{\prime}_{f,t} \in \mathbb{R}^{P \times P}$ and $\bar{\ma{b}}_{f,t} \in \mathbb{R}^{P \times 1}$ are the forget gate weights and biases at time $t$, $\bar{\ma{x}}_{t} \in \mathbb{R}^{K_{in} \times 1}$ and $\bar{\ma{z}}_{t-1}$ represents the {\ac{LSTM}} unit input vector of size $K_{in}$, and the previous hidden state of size $P$, respectively.

\paragraph{Store the relevant new information} After classifying the relevant information, the {\ac{LSTM}} unit applies some computations on the selected information via the input gate
    \begin{equation}
    {\bar{\ma{i}}_{t}} = \sigma (\ma{W}_{\bar{\ma{i}}, t}\bar{\ma{x}}_{t} + \ma{W}^{\prime}_{\bar{\ma{i}},t}\bar{\ma{z}}_{t-1} + \bar{\ma{b}}_{\bar{\ma{i}},t}),
    \label{eq: lstm_ing}
    \end{equation}
    \begin{equation}
    {\tilde{{\ma{c}}}}_{t} = \text{tanh} (\ma{W}_{{\tilde{{\ma{c}}}}, t}\bar{\ma{x}}_{t} + \ma{W}^{\prime}_{{\tilde{{\ma{c}}}},t}\bar{\ma{z}}_{t-1} + \bar{\ma{b}}_{{\tilde{{\ma{c}}}},t}).
    \label{eq: lstm_incg}
    \end{equation}
    
\paragraph{Update the new cell state} Next the {\ac{LSTM}} unit is supposed to update the current cell state ${{{\ma{c}}}}_{t}$ based on the two previously-mentioned steps such that
    \begin{equation}
    {{{\ma{c}}}}_{t} = {\ma{f}}_{t} \odot {\ma{c}}_{t-1} +  \bar{\ma{i}}_{t} \odot {\tilde{{\ma{c}}}}_{t}. 
    \label{eq: lstm_cell_state}
    \end{equation}
    where $\odot$ denotes the Hadamard product. 
    \paragraph{Generate the LSTM unit output} Updating the hidden state and generating the output by the output gate is the final processing step. The output is considered to be a cell state filtered version and can be computed such that
    \begin{equation}
    {\ma{o}}_{t} = \sigma (\ma{W}_{o, t}\bar{\ma{x}}_{t} + \ma{W}^{\prime}_{o,t}\bar{\ma{z}}_{t-1} + \bar{\ma{b}}_{o,t}),
    \label{eq: lstm_og}
    \end{equation}
    \begin{equation}
    {\bar{{\ma{z}}}}_{t} =  {\ma{o}}_{t} \odot \text{tanh}({\ma{c}}_{t}).
    \label{eq: lstm_hidden_state}
    \end{equation}

In literature, there exists several {\ac{LSTM}} architecture variants, where the interactions between the {\ac{LSTM}} unit gates are modified. The authors in~\cite{ref_lstm_var} provide a detailed comparison of popular {\ac{LSTM}} architecture variants.

\subsection{CNN}
Another type of deep learning is  \ac{CNN} model. This is commonly used for processing data with grid patterns, such as images~\cite{ref_CNN}. Thus, {\ac{CNN}} has generally become the state of the art for several visual applications such as image classification, due to its demonstrated ability to extract patterns from the input image. {\ac{CNN}} can be seen as a set of several layers stacked together to accomplish the requisite task. These layers include
\begin{itemize}
    \item Input layer: It represents the 2D or 3D input image. For the sake of simplicity, let us consider a 2D image input  to the $l$ -th {\ac{CNN}} layer denoted by $\ma{X}_{l} \in \mathbb{R}^{h_{l} \times w_{l}}$, where $h_{l}$ and $w_{l}$ denote the height and the width of the $\ma{X}_{l}$ input image.
    
    \item Convolutional layer: refers to a specialized type of linear operation used for feature extraction, where predefined filters referred to as kernels scan the input matrix to fill the output matrix denoted as feature map, which is shown in \figref{fig:cnn_conv_ex}. We note that different kernels can be considered as different feature extractors. 
    
    Two key hyper parameters define the \ac{CNN} convolutional layer, namely, the  size and number of kernels denoted by $f_{l}$ and $n_{l}$, respectively. The typical kernel size is $3\times3$, but sometimes $5\times5$ or $7\times7$. The number of kernels is arbitrary and determines the depth of output feature maps. It is possible to tune these parameters according to the application type. Furthermore, the process of training a \ac{CNN} model regarding the convolution layer involves identifying the kernels values that work optimally for a particular task based on a given training dataset. In the convolution layer, the kernels are the only automatically learned parameters during the training process. Mathematically speaking, for a given input image $\ma{X}_{l}$ and kernel $\ma{K_{l}} \in  \mathbb{R}^{f_{l} \times f_{l} \times 1}$, we consider one kernel for simplicity, the generated feature map $\ma{Y}_{l} \in  \mathbb{R}^{ (h_{l} - f + 1) \times (w_{l} - f + 1)}$ can be expressed as follows
    \begin{equation}
        \ma{Y}_{l} [x, y] = \sum^{h_{l}}_{i = 1} \sum^{w_{l}}_{j = 1}  \ma{K}_{l} [i,j] \ma{X}_{l} [ x + i - 1, y + j - 1].
    \end{equation}
    
\begin{figure}[t]
\centering
\includegraphics[width=0.7\columnwidth]{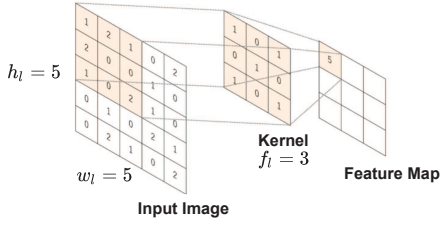}
\caption{CNN convolutional layer example~\cite{gizzini_thesis}.}
\label{fig:cnn_conv_ex}
\end{figure}

        \item Activation layer: The outputs of a linear operation such as convolution pass through a nonlinear activation function. This activation function introduces non-linear processing to the {\ac{CNN}} architecture given that the input-output {\ac{CNN}} pairs relation could be non-linear. While several non-linear activation functions exist such as sigmoid or hyperbolic tangent (tanh) function, the most common presently used function is the rectified linear unit (ReLU).   
    
        \item Pooling layer: This layer is employed to decrease the number of parameters when the images are too large. Pooling operation is also referred to as sub-sampling or down-sampling. This reduces the dimensionality of all feature maps but does manage to retain significant information. Notably, none of the pooling layers contains any learnable parameter. The most popular form of pooling operation is max pooling, which extracts patches from the input feature maps, outputs the maximum value in each patch, and then discards all the other values. However, there are other pooling operations such as global average pooling~\cite{ref_cnn_avg_pooling}. 
    
    \begin{figure}[t]
\centering
\includegraphics[width=1\columnwidth]{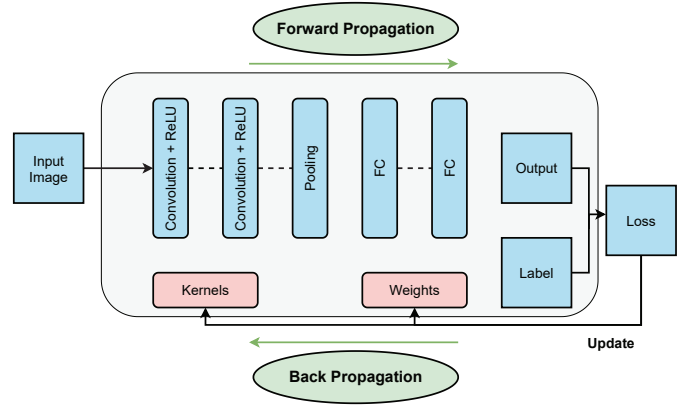}
\caption{CNN classical architecture~\cite{gizzini_thesis}.}
\label{fig:cnn_archi}
\end{figure}
    \item Fully connected layer: This layer forms the last block of the {\ac{CNN}} architecture and is mainly employed in the classification problems. It is a simple feed-forward neural network layer that comprises at least one hidden layer; its role is to transform the 2D \ac{CNN} layer output into a 1D vector. In classification problems, the final outputs of the \ac{CNN} network represent the probabilities for each class, where the final fully-connected layer typically has the same number of output nodes as the number of classes.
    \item Batch normalization: It is used to increase the {\ac{CNN}} stability of the output by normalizing each layer's output. Moreover, batch normalization layer reduces overfitting and accelerates the {\ac{CNN}} training.
    \item Output layer: This layer is configured in accordance with the studied problem. For instance, in classification problems the {\ac{CNN}} output layer is a fully connected layer with softmax activation function. On the other hand, in regression problems, the {\ac{CNN}} output does not use any activation function.
\end{itemize}

Figure~\ref{fig:cnn_archi} illustrates the classical CNN architecture. As seen in this figure, the only trainable parameters within the \ac{CNN} network are the kernels and the fully connected layer weights. Similar to all other {\ac{DL}} techniques, {\ac{CNN}} network updates its trainable parameters by minimizing the {\ac{CNN}} loss function that measures how far the inputs are from the outputs. Thereafter, the \ac{CNN} kernels and weights are updated in the back propagation operation~\cite{cnn_backpro}. Finally, the performance of the trained {\ac{CNN}} model is examined in the testing phase where new unobserved images are fed to the trained {\ac{CNN}} model.

It is noteworthy that there are special {\ac{CNN}} architectures such as {\ac{SR-CNN}}~\cite{ref_SRCNN}, {\ac{DN-CNN}}~\cite{ref_DNCNN}, and {\ac{SR-ConvLSTM}}~\cite{ref_ConvLSTM_Comp} that are mainly used for regression problems. {\ac{SR-CNN}} is used for enhancing the quality of the input image, where it takes the low-resolution image as the input and outputs the high-resolution one. \ac{DN-CNN} uses another methodology to improve the image quality by separating the noise from the input noisy image employing  residual learning~\cite{ref_residual_learning}. The input noisy image is then subtracted from the extracted noise, resulting in the denoised image. Furthermore, {\ac{SR-ConvLSTM}} combines both {\ac{LSTM}} and {\ac{CNN}} networks together where time correlation across the whole input image is learned, thus leading to a better estimation accuracy.
\section{DL-Based SBS Channel Estimation} 
\label{sbs_estimators}


In {\ac{DL}}-based {\ac{SBS}} channel estimation, \ac{FNN} and {\ac{LSTM}} networks are primarily integrated with conventional estimation schemes in the following two manners:
(\textit{i}) \ac{FNN} is implemented as a post-processing module after conventional {\ac{DPA}}, {\ac{STA}}, and {\ac{TRFI}} estimators. (\textit{ii}) {\ac{LSTM}} network gets implemented as a pre-processing unit before conventional {\ac{DPA}} estimation to minimize the DPA demapping error iteratively. Both implementations are helpful in improving the channel estimation's accuracy, particularly in high mobility scenarios. However, the LSTM-based estimation illustrates a considerable superiority over the \ac{FNN}-based estimation as demonstrated in Section~\ref{simulation_results}. Hereafter, the steps applied in each DL-based {\ac{SBS}} estimator are presented.

\subsection{DPA-FNN} \label{SBS_AE_DNN}

\begin{figure*}[t]
\centering
\includegraphics[width=2\columnwidth]{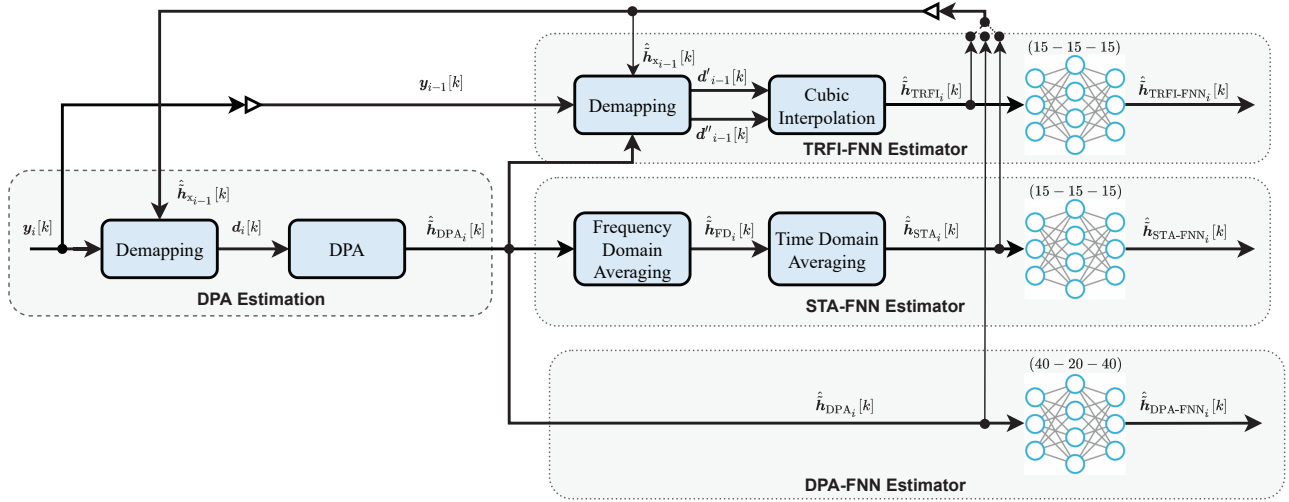}
\caption{The block diagram of the studied DNN-based SBS estimators.}\label{fig:DNN_based_Estimators}
\end{figure*}

The {\ac{DPA}} estimation~\cite{ref_sta} utilizes the demapped data subcarriers of the previously received {\ac{OFDM}} symbol for estimating the channel for the existing {\ac{OFDM}} symbol such that

\begin{equation}
\tilde{\ma{d}}_i[k] =  \mathfrak{D} \big( \frac{\tilde{\ma{y}}_i[k]}{\hat{\tilde{\ma{h}}}_{\text{DPA}_{i-1}}[k]}\big)
,~ \hat{\tilde{\ma{h}}}_{\text{DPA}_{0}}[k] = \hat{\tilde{\ma{h}}}_{\text{LS}}[k], 
\label{eq: DPA_1}
\end{equation}
where $\mathfrak{D}(.)$ refers to the demapping operation to the nearest constellation point in accordance with the employed modulation order. $\hat{\tilde{\ma{h}}}_{\text{LS}}$ signifies the LS estimated channel at the received preambles, such that
\begin{equation}
\hat{\tilde{\ma{h}}}_{\text{LS}}[k] = \frac{\sum\limits_{u=1}^{P}\tilde{\ma{y}}^{(p)}_{u}[k]}{P\tilde{\ma{\Lambda}}[k]},~k \in \Kon,
\label{eq: LS}
\end{equation}
where $\tilde{\ma{\Lambda}}$ denotes the frequency domain predefined preamble sequence. Thereafter, the final {\ac{DPA}} channel estimates are updated in the following manner 
\begin{equation}
\hat{\tilde{\ma{h}}}_{\text{DPA}_{i}}[k] = \frac{\tilde{\ma{y}}_i[k]}{\tilde{\ma{d}}_i[k]}.
\label{eq: DPA_2}
\end{equation}

{\ac{DPA}} estimation suffers from two main limitations. First, it is based on the basic $\hat{\tilde{\ma{h}}}_{\text{LS}}$ estimation suffering from noise enhancement. Second, the demapping step in {\ac{DPA}} leads to a significant demapping error primarily in low {\ac{SNR}} region stemming from the noise imperfections and doubly-dispersive channel variations. This demapping error is enlarged in high mobility scenarios employing high modulation orders. In addition, since the {\ac{DPA}} estimated channels are updated iteratively over the received frame, the demapping error propagates via the frame that results in a significant  degradation in performance. In order to address these limitations, the DPA-{\ac{FNN}} scheme~{\cite{ref_dpa_dnn}} has been proposed to compensate the {\ac{DPA}} estimation error, where $\hat{\tilde{\ma{h}}}_{\text{DPA}_{i}}[k]$ is fed to a three-hidden-layers \ac{FNN} with $40-20-40$ neurons, as shown in \figref{fig:DNN_based_Estimators}. Using the \ac{FNN} in addition to the {\ac{DPA}} scheme yields good performance but it is not sufficient, because it ignores the time and frequency correlation between successive received {\ac{OFDM}} symbols. Also, the employed \ac{FNN} architecture can be optimized to reduce the  computational complexity of channel estimation.
\subsection{STA-FNN}
To improve the conventional \ac{DPA} estimation, the {\ac{STA}} estimator~\cite{ref_sta} has been proposed where frequency and time-domain averaging are applied on top of the {\ac{DPA}} estimated channel as follows
\begin{equation}
\hat{\tilde{\ma{h}}}_{\text{FD}_{i}}[k] = \sum_{\lambda = -\beta}^{\lambda = \beta} \omega_{\lambda} \hat{\tilde{\ma{h}}}_{\text{DPA}_{i}}[k + \lambda], ~ \omega_{\lambda} = \frac{1}{2\beta+1}.
\label{eq: STA_4}
\end{equation}
\begin{equation}
\hat{\tilde{\ma{h}}}_{\text{STA}_{i}}[k] = (1 - \frac{1}{\alpha})  \hat{\tilde{\ma{h}}}_{\text{STA}_{i-1}}[k] + \frac{1}{\alpha}\hat{\tilde{\ma{h}}}_{\text{FD}_{i}}[k].
\label{eq: STA_5}
\end{equation}
{\ac{STA}} estimator performs well in the low {\ac{SNR}} region. However, it suffers from a considerable error floor in high {\ac{SNR}} regions due to the large {\ac{DPA}} demapping error. Importantly, in~\cite{ref_sta}, the values of the frequency and time averaging coefficients are fixed to $\alpha = \beta = 2$. Thus, the final STA estimated channel is a linear combination between the previously estimated channel~{\eqref{eq: STA_5}} and the frequency averaged channel estimates~{\eqref{eq: STA_4}}. However, this linear combination leads to a significant performance degradation in real case scenarios due to the doubly-dispersive channel non-linear imperfections. Here, {\ac{FNN}} is utilized as a post non-linear processing unit after the conventional {\ac{STA}} scheme~\cite{ref_STA_DNN}. STA-FNN captures more the time-frequency correlations of the channel samples, apart from correcting the conventional {\ac{STA}} estimation error. Furthermore, the optimized STA-FNN architecture performs better than the DPA-FNN with a significant computational complexity decrease, as elucidated in Section~\ref{complexity}.
\subsection{TRFI-FNN}
{\ac{TRFI}} estimation scheme~\cite{ref_trfii} is another methodology used for improving the {\ac{DPA}} estimation in~\eqref{eq: DPA_2}. Assuming that the time correlation of the channel response between two adjacent {\ac{OFDM}} symbols is high, {\ac{TRFI}} define two sets of subcarriers such that: (\textit{i}) {$\RS_{i}$} set: that includes the reliable subcarriers indices,  and (\textit{ii}) {$\URS_{i}$} set: which contains the unreliable subcarriers indices. The estimated channels for the {$\URS_{i}$} are then interpolated using the {$\RS_{i}$} channel estimates by means of the frequency-domain cubic interpolation. This procedure can be expressed in the following manner
\begin{itemize}
    \item Equalize the previously received {\ac{OFDM}} symbol by ${\hat{\tilde{\ma{h}}}_{\text{TRFI}_{i-1}}[k]}$ and ${\hat{\tilde{\ma{h}}}_{\text{DPA}_{i}}[k]}$, such that
    \begin{equation}
    \begin{split}
    {\tilde{\ma{d}}^\prime}_{i-1}[k] = \mathfrak{D} \big( \frac{\tilde{\ma{y}}_{i-1}[k]}{\hat{\tilde{\ma{h}}}_{\text{DPA}_{i}}[k]} \big), ~
    {\tilde{\ma{d}}^{\prime\prime}}_{i-1}[k] =  \mathfrak{D} \big( \frac{\tilde{\ma{y}}_{i-1}[k]}{\hat{\tilde{\ma{h}}}_{\text{TRFI}_{i-1}}[k]} \big). 
    \label{eq: TRFI_1}
    \end{split}
    \end{equation}
    \item According to the demapping results, the subcarriers are grouped as follows
    \begin{equation}
       \left\{
        \begin{array}{ll}
            \RS_{i} \leftarrow \RS_{i} + {k} ,&\quad \tilde{\ma{d^\prime}}_{i-1}[k] = \tilde{\ma{d}}^{\prime\prime}_{i-1}[k] \\
            \URS_{i} \leftarrow \URS_{i} + {k},&\quad \tilde{\ma{d^\prime}}_{i-1}[k] \neq \tilde{\ma{d}}^{\prime\prime}_{i-1}[k]
        \end{array}\right. .
       \label{eq: TRFI_2}
       \end{equation}
\item Finally, frequency-domain cubic interpolation is employed to estimate the channels at the {$\URS_{i}$} as follows
\begin{equation}
      \hat{\tilde{\ma{h}}}_{\text{TRFI}_{i}}[k] = \left\{
        \begin{array}{ll}
            \hat{\tilde{\ma{h}}}_{\text{DPA}_{i}}[k] ,&\quad k \in \RS_{i} \\
            \text{Cubic Interpolation},&\quad k \in \URS_{i}
        \end{array}\right. .
       \label{eq: TRFI_3}
       \end{equation}
\end{itemize}
Performing frequency-domain interpolation in addition to the {\ac{DPA}} estimation enhances the performance. However, {\ac{TRFI}} still suffers from the demapping and interpolation errors as the number of \ac{RS} subcarriers is inversely proportional to the channel variations. Additionally, the condition where ${\tilde{\ma{d}}^{\prime}}_{i-1}[k] \neq {\tilde{\ma{d}}^{\prime\prime}}_{i-1}[k]$ is more dominant in high mobility scenarios. It is for this reason that only a few \ac{RS} subcarriers will be selected and the employed cubic interpolation performance will be degraded.

Inspired by the work undertaken in {\ac{STA}}-{\ac{FNN}}, the authors in~\cite{ref_TRFI_DNN} used the same optimized \ac{FNN} architecture as in~\cite{ref_STA_DNN}, albeit with $\hat{\tilde{\ma{h}}}_{\text{TRFI}_{i}}[k]$ as an input instead of  $\hat{\tilde{\ma{h}}}_{\text{STA}_{i}}[k]$. {\ac{TRFI}}-{\ac{FNN}} corrects the cubic interpolation error and also learns the channel frequency domain correlation, thus leading to an improved performance in high {\ac{SNR}} regions.

\subsection{LSTM-FNN-DPA}

\begin{figure*}[t]
	\setlength{\abovecaptionskip}{6pt plus 3pt minus 2pt}
	\centering
  \includegraphics[width=0.95\textwidth]{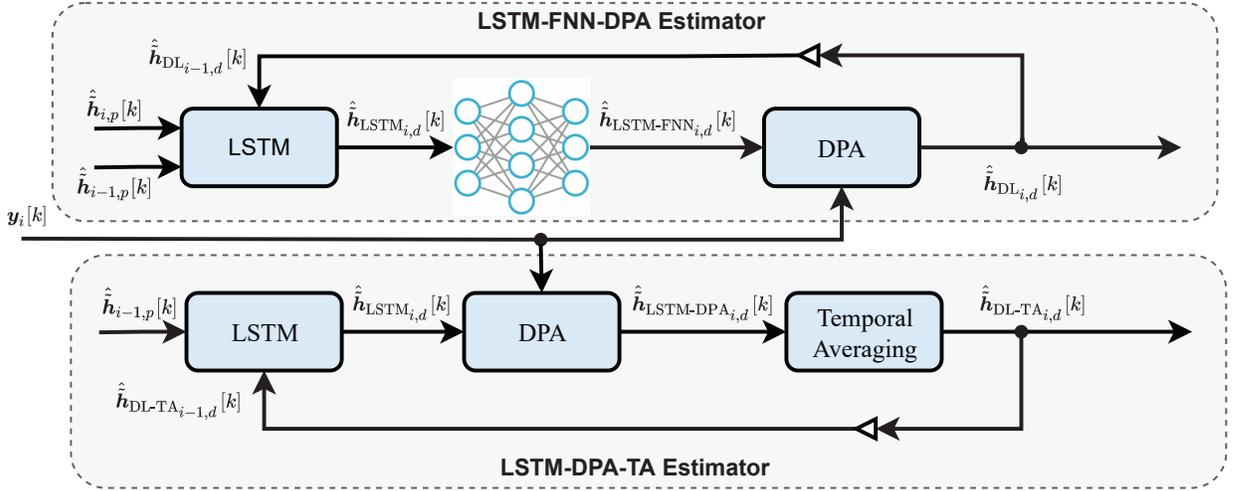}
  \caption{The block diagram of the studied LSTM-based SBS estimators.}
  \label{fig:block_diagram}
\end{figure*}

Unlike the \ac{FNN}-based estimators, where the {\ac{DL}} processing is employed following the conventional estimators, the work carried out in~\cite{ref_lstm_dnn_dpa} shows that employing the {\ac{DL}} processing prior to the conventional estimator, specifically the {\ac{DPA}} estimation, could lead to a significant improvement in the overall performance. In this context, the authors have proposed to use two cascaded {\ac{LSTM}} and \ac{FNN} networks for both channel estimation as well as noise compensation, as shown in \figref{fig:block_diagram}. 

The LSTM-FNN-DPA estimator employs the previous and current pilot subcarriers besides the \ac{LSTM}-{\ac{FNN}} estimated channel employed in the {\ac{DPA}} estimation, such that
\begin{equation}
    \tilde{\ma{d}}_{\text{LSTM-FNN}_{i,d}}[k] =  \mathfrak{D} \big( \frac{\tilde{\ma{y}}_{i,d}[k]}{\hat{\tilde{\ma{h}}}_{\text{LSTM-FNN}_{i-1,d}}[k]}\big)
    ,~ \hat{\tilde{\ma{h}}}_{\text{LSTM}_{0}}[k] = \hat{\tilde{\ma{h}}}_{\text{LS}}[k],
    \label{eq: proposed3}
    \end{equation}
    \begin{equation}
    \hat{\tilde{\ma{h}}}_{\text{DL}_{i,d}}[k] = \frac{\tilde{\ma{y}}_{i,d}[k]}{\ma{d}_{\text{LSTM}_{i,d}}[k]}.
    \label{eq: proposed44}
    \end{equation}
    While this estimator can outperform the  \ac{FNN}-based estimators, it experiences a considerable computational complexity arising from the employment of two {\ac{DL}} networks.
\subsection{LSTM-DPA-TA}
The authors in~\cite{gizzini:hal-03365697} propose to use only LSTM network instead of two as implemented in the LSTM-FNN-DPA estimator. In addition, noise compensation is made possible by applying {\ac{TA}} processing as shown in \figref{fig:block_diagram}. This methodology only requires the previous pilots besides the LSTM estimated channel as an input.
Then, the LSTM estimated channel is employed in the DPA estimation as follows
    \begin{equation}
    \tilde{\ma{d}}_{\text{LSTM}_{i}}[k] =  \mathfrak{D} \big( \frac{\tilde{\ma{y}}_i[k]}{\hat{\tilde{\ma{h}}}_{\text{LSTM}_{i-1}}[k]}\big)
    ,~ \hat{\tilde{\ma{h}}}_{\text{LSTM}_{0}}[k] = \hat{\tilde{\ma{h}}}_{\text{LS}}[k],
    \label{eq: proposed33}
    \end{equation}
    \begin{equation}
    \hat{\tilde{\ma{h}}}_{\text{LSTM-DPA}_{i}}[k] = \frac{\tilde{\ma{y}}_i[k]}{\tilde{\ma{d}}_{\text{LSTM}_{i}}[k]}.
    \label{eq: proposed4}
    \end{equation}
Finally, to alleviate the impact of the AWGN noise, {\ac{TA}} processing is applied to the $\hat{\tilde{\ma{h}}}_{\text{LSTM-DPA}_{i}}[k]$ estimated channel, such that
    \begin{equation}
    \hat{\bar{\ma{h}}}_{\text{DL-TA}_{i,d}} = (1 - \frac{1}{\alpha})  \hat{\bar{\ma{h}}}_{\text{DL-TA}_{i - 1,d}} + \frac{1}{\alpha}  \hat{\bar{\ma{h}}}_{\text{LSTM-DPA}_{i,d}}.
     \label{eq: proposed5}
\end{equation}
Here, $\alpha$ denotes the utilized weighting coefficient. In~\cite{gizzini:hal-03365697}, the authors use a fixed $\alpha = 2$ for simplicity. Therefore, the {\ac{TA}} applied in~{\eqref{eq: proposed5}} reduces the AWGN noise power $\sigma^2$ iteratively within the received {\ac{OFDM}} frame according to the ratio
\begin{equation}
	\begin{split}
			{R}_{\text{DL-TA}_{q}} &= \left( \frac{1}{4} \right)^{(q-1)} + \sum_{j=2}^{q} \left( \frac{1}{4} \right)^{(q-j+1)}=\frac{4^{q-1} + 2}{3 \times 4^{q-1}}.
	\label{eq:noise_degradtion}
	\end{split}
\end{equation}
This corresponds to the AWGN noise power ratio of the estimated channel at the $q$-th estimated channel, where ${1 < q < I + 1}$ and ${{R}_{\text{DL-TA}_{1}} = 1}$ denotes the AWGN noise power ratio at $\hat{\tilde{\ma{h}}}_{\text{LS}}[k]$. From the derivation of ${R}_{\text{DL-TA}_{q}}$, it can be seen that the noise power decreases over the received {\ac{OFDM}} frame, i.e., the SNR increases, resulting in an overall improved performance. Moreover, the input dimension reduction, coupled with the simple {\ac{TA}} processing, significantly lowers the overall computational complexity.
\begin{table}[ht]
    \small
	\renewcommand{\arraystretch}{1.3}
	\centering
	\caption{Parameters of the studied DL-based SBS channel estimators.}
	\label{tb:DNN_LSTM_params}
	\begin{tabular}{l|l}
				\hline
		DPA-\ac{FNN} (Hidden layers; Neurons per layer) & (3;40-20-40)  \\ \hline
		STA-\ac{FNN} (Hidden layers; Neurons per layer) & (3;15-15-15)  \\ \hline
		TRFI-\ac{FNN} (Hidden layers; Neurons per layer) & (3;15-15-15)  \\ \hline
	    LSTM (Hidden layers; Neurons per layer) & (1;128)  \\ \hline
		
		Activation function              & ReLU                     \\ \hline
		Number of epochs        & 500                                \\ \hline
		Training samples        & 800000                             \\ \hline
		Testing samples        & 200000                             \\ \hline
		Batch size          & 128                                    \\ \hline
		Optimizer       & ADAM                                       \\ \hline
		Loss function      & MSE                                     \\ \hline
		Learning rate        & 0.001                                 \\ \hline
	    Training SNR        & 40 dB                                 \\ \hline
	\end{tabular}
\end{table}
Intensive experiments reveal that the performance of DL networks is strongly related to the SNR considered in the training~{\cite{r20}}. The training undertaken at the highest SNR value provides the best performance. In fact, the DL network is able to learn better the channel when the training is performed at a high SNR value because the impact of the channel is higher than the impact of the noise in this SNR range. Owing to the robust generalization properties of DL, trained networks can still estimate the channel even if the noise increases, i.e., at low SNR values. Therefore, {\ac{FNN}} and LSTM based estimators training is performed using {\ac{SNR}} = $40$ dB to attain the best performance. Moreover, intensive experiments are performed using the grid search algorithm~\cite{ref_grid} to select the most suitable \ac{FNN} and LSTM hyper parameters in terms of performance as well as complexity. Figures~\ref{fig:DNN_based_Estimators} and~\ref{fig:block_diagram} illustrate the block diagram of the \ac{FNN} and LSTM based estimators. Furthermore, Table~\ref{tb:DNN_LSTM_params} presents their parameters. 

\section{DL-Based FBF Channel Estimation Schemes} 
\label{fbf_estimators}
This section presents the \ac{DL}-based {\ac{FBF}} estimators introduced to improve the channel estimation accuracy, particularly in very high mobility scenarios, where the channel variation is found to be severe. Similar to the {\ac{DL}}-based {\ac{SBS}} estimators, the \ac{DL}-based {\ac{FBF}} estimators apply first conventional estimation followed by means of {\ac{CNN}} processing. 
%
%
%
\begin{table*}[]
\small
\renewcommand{\arraystretch}{1}
\centering
\caption{Main characteristics and features of the studied DL-based channel estimators.}
\label{tb:SoA_Summary}
\begin{tabular}{|c|c|c|c|c|c|c|l|}
\hline
\begin{tabular}[c]{@{}c@{}}Estimator \\ type\end{tabular} & \begin{tabular}[c]{@{}c@{}}Estimator \\ reference\end{tabular} & \begin{tabular}[c]{@{}c@{}}Conventional \\ estimation\end{tabular} & \begin{tabular}[c]{@{}c@{}}DL-based \\ Method\end{tabular}   & Complexity & \begin{tabular}[c]{@{}c@{}}BER \\ Performance\end{tabular} & Robustness & \multicolumn{1}{c|}{Pros and Cons}                                                                                                                                                                                                                                                                           \\ \hline
\multirow{5}{*}{SBS}                                      & \cite{ref_dpa_dnn}                                                              & DPA                                                                & \multirow{3}{*}{FNN}                                         & ++         & ++                                                         & ++         & \begin{tabular}[c]{@{}l@{}}+ Significant performance \\ superiority over\\ conventional estimators. \\ - Ignore the time and \\ frequency correlation \\ between successive \\ received OFDM symbols.\\ - Complex FNN architecture \\ to compensate the \\ conventional DPA \\ demapping error.\end{tabular} \\ \cline{2-3} \cline{5-8} 
                                                          & \cite{ref_STA_DNN}                                                              & STA                                                                &                                                              & +          & +++                                                        & ++         & \begin{tabular}[c]{@{}l@{}}+ STA averaging ameliorate the\\ impact of the AWGN noise\\  in low SNR regions.\\ + Optimized FNN architecture.\\ - Fixed averaging coefficients.\\ - Performance degradation in\\  high mobility scenarios.\end{tabular}                                                         \\ \cline{2-3} \cline{5-8} 
                                                          & \cite{ref_TRFI_DNN}                                                            & TRFI                                                               &                                                              & +          & ++++                                                       & +++        & \begin{tabular}[c]{@{}l@{}}+ Cubic Interpolation enhances \\ the performance in the entire \\ SNR region.\\ + Optimized FNN architecture.\\ - Assume high correlation \\ between successive OFDM \\ symbols.\\ - Lack of robustness in very \\ high mobiliy scenarios.\end{tabular}                            \\ \cline{2-8} 
                                                          & \cite{ref_lstm_dnn_dpa}                                                            & DPA                                                                & LSTM and FNN                                                 & +++        & ++++                                                       & ++++       & \begin{tabular}[c]{@{}l@{}}+ Outperform FNN-based \\ estimators.\\ + Improved estimation \\ since LSTM is implemented \\ before DPA estimation\\ - Employ LSTM and \\ FNN in the same architecture.\end{tabular}                                                                                             \\ \cline{2-8} 
                                                          & \cite{gizzini:hal-03365697}                                                              & DPA and TA                                                         & LSTM                                                         & +++        & ++++                                                       & ++++       & \begin{tabular}[c]{@{}l@{}}+ TA processing results in\\ a considerable decline in\\ the AWGN noise.\\ + Employ only one \\ optimized LSTM unit.\\ + Reduced input dimension.\end{tabular}                                                                                                                     \\ \hline
\multirow{3}{*}{FBF}                                      & \cite{ref_ChannelNet}                                                             & 2D RBF                                                             & \begin{tabular}[c]{@{}c@{}}SR-CNN  and\\ DN-CNN\end{tabular} & +++++      & ++                                                         & ++         & \begin{tabular}[c]{@{}l@{}}- 2D RBF interpolation high\\ computational complexity.\\ - The 2D RBF function \\ and scale factor should be\\ optimized in accordance with \\ the channel variations.\\ - Employ two high-complexity\\ CNN architectures.\end{tabular}                                                \\ \cline{2-8} 
                                                          & \cite{ref_TS_ChannelNet}                                                              & ADD-TT                                                             & SR-ConvLSTM                                                  & +++++      & +++                                                        & +++        & \begin{tabular}[c]{@{}l@{}}+ Outperform ChannelNet\\ estimator {[}29{]}.\\ - Fixed ADD-TT Averaging \\ coefficients.\\ - High computational complexity\\ owing to the integration of both\\ LSTM and CNN architectures.\end{tabular}                                                                          \\ \cline{2-8} 
                                                          & \cite{ref_WI_CNN}                                                             & WI                                                                 & \begin{tabular}[c]{@{}c@{}}SR-CNN or \\ DN-CNN\end{tabular}  & +++        & ++++                                                       & ++++       & \begin{tabular}[c]{@{}l@{}}+ Adaptive frame structure \\ according to the mobility\\  condition.\\ + Reduced buffering time \\ at the receiver.\\ + Transmission data rate gain.\\ + Optimized CNN \\ architectures.\end{tabular}                                                                            \\ \hline
\end{tabular}
\end{table*}
\subsection{ChannelNet}
In~\cite{ref_ChannelNet}, the authors use forward a \ac{CNN}-based channel estimator denoted as {\ac{ChannelNet}} scheme, where 2D \ac{RBF} interpolation is implemented as an initial channel estimation. The underlying motivation of the 2D {\ac{RBF}} interpolation is to approximate multidimensional scattered unknown data from their surrounding neighbors known data by employing the radial basis function. In order to achieve the purpose, the distance function is calculated between every data point to be interpolated and its neighbours, where closer neighbors are assigned higher weights. Thereby, the \ac{RBF} interpolated frame is considered a low resolution image, where {\ac{SR-CNN}} is utilized to obtain an improved estimation. Finally, to ameliorate the effect of noise within the high resolution estimated frame, \ac{DN-CNN} is implemented leading to a high resolution and noise alleviated estimated channels. The ChannelNet estimator considers sparsely allocated pilots within the IEEE 802.11p frame and initially applies the {\ac{LS}} estimation to the pilot subcarriers within the received {\ac{OFDM}} frame. Subsequently, the 2D {\ac{RBF}} interpolation is derived by the weighted summation of the distance between each data subcarrier to be interpolated as well as all the pilot subcarriers in the received {\ac{OFDM}} frame, such that
\begin{equation}
\hat{\tilde{\ma{H}}}_{\text{RBF}}[k,i] = \sum_{j = 1}^{{K_{{p}}}I} \omega_j \Phi (| k - \Kf[j]| ,  |i - \Kt[j]|).
\label{eq:RBF_eq}    
\end{equation}

$\Kf = [{\set{K}}_{\text{p}_{1}}, \dots, {\set{K}}_{\text{p}_{I}}] \in \mathbb{R}^{1 \times K_{p}I} $ and $\Kt = [(1)_{\times K_{p}},\dots, (I)_{\times K_{p}}]  \in \mathbb{R}^{1 \times K_{p}I}$ represent the frequency and time indices vectors of the  allocated pilot subcarriers within the received {\ac{OFDM}} frame, respectively.  $\omega_j$ is the {\ac{RBF}} weight multiplied by the {\ac{RBF}} interpolation function $\Phi(.)$ between the $(k, i)$ data subcarrier and the $(\Kf[j], \Kt[j])$ pilot subcarrier. In~\cite{ref_ChannelNet}, the {\ac{RBF}} gaussian function is applied, such that
\begin{equation}
    \Phi(x,y) = e^{-\frac{(x + y)^2}{r_{0}}}. \label{eq:RBF_PHI}
\end{equation}

$r_{0}$ refers to the 2D {\ac{RBF}} scale factor that varies based on the used {\ac{RBF}} function. Notably, altering the value of $r_{0}$ alters the shape of the interpolation function. Moreover, the \ac{RBF} weights $\ma{w}_{\text{RBF}} = [\omega_{1}, \dots, \omega_{K_{p}I}] \in \mathbb{R}^{K_{p}I \times 1}$ are calculated using the following relation:
\begin{equation}
    \ma{A}_{\text{RBF}} \ma{w}_{\text{RBF}} = \bar{{\ma{h}}}_{\text{LS}}.
\label{eq:RBF_w}
\end{equation}

Here, $\ma{A}_{\text{RBF}} \in \mathbb{R}^{ {K_{{p}}}I  \times {K_{{p}}}I }$ is the {\ac{RBF}} interpolation matrix  of the pilots subcarriers, with entries $a_{i,j} = \Phi(\Kf[i],\Kt[j])$ where $i,j = 1, \dots, K_{p}I$. It is observed that, $\bar{{\ma{h}}}_{\text{LS}} = \Vect{\hat{\tilde{\ma{H}}}_{\text{LS}}} \in \compl ^{K_{p}I \times 1}$ is a vector that contains the {\ac{LS}} estimated channels at all the pilot subcarriers within the received {\ac{OFDM}} frame. This is expressed as
\begin{equation}
\hat{\tilde{\ma{H}}}_{\text{LS}}[k,i] = \frac{\tilde{\ma{Y}}[k,i]}{\tilde{\ma{P}}[k,i]},~ k \in {\set{K}}_{\text{p}},~ 1\leq i \leq I,
\label{eq: LS-RBF}
\end{equation}
with $\tilde{\ma{P}}[k,i]$ is the frequency-domain pre-defined pilot subcarriers, and ${\set{K}}_{\text{p}}$ refers to the allocated sparse pilots indices within the received {\ac{OFDM}} symbol.
After computing $\ma{W}_{\text{RBF}}$, it is possible to calculate the {\ac{RBF}} estimated channel for every data subcarriers within the received {\ac{OFDM}} frame, as shown in~\eqref{eq:RBF_eq}. Finally, the {\ac{RBF}} interpolation estimated frame $\hat{\tilde{\ma{H}}}_{\text{RBF}}$ is fed as an input to {\ac{SR-CNN}} and {\ac{DN-CNN}} to improve the channel estimation accuracy and reduce the noise impact.

The ChannelNet estimator limitations lie in: (\textit{i}) 2D {\ac{RBF}} interpolation high computational complexity arising from the computation of~{\eqref{eq:RBF_w}} for the channel estimation of all data subcarriers. (\textit{ii}) The 2D {\ac{RBF}} function and scale factor needs to be optimized in accordance with the channel variations. (\textit{iii}) The integrated {\ac{SR-CNN}} and {\ac{DN-CNN}} architectures have significant computational complexity. Notably, the ChannelNet estimator uses a fixed  {\ac{RBF}} function and scale factor, thus experiencing a considerable degradation in performance, particularly in low {\ac{SNR}} regions, where the noise impact remains dominant, as well as high mobility vehicular scenarios, where the channel varies swiftly within the {\ac{OFDM}} frame.
\begin{table}[t]
    \small
	\renewcommand{\arraystretch}{1.3}
	\centering
	\caption{Parameters of the studied DL-based FBF channel estimators.}
	\label{tb:CNN_params}
	\begin{tabular}{l|l}
		\hline
		\textbf{Parameter}      & \textbf{Values}                    \\ \hline
		Input/Output dimensions & $ 2K_{\text{on}} \times I \times 1$   \\ \hline
		SR-CNN (Hidden layers - $n_{l},f_{l}$) & (3 - 9,64;~1,32;~ 5,1)   \\ \hline
		DN-CNN (Hidden layers -  $n_{l},f_{l}$) & (18 - 64,~3)   \\ \hline
		Optimized SR-CNN (Hidden layers - $n_{l},f_{l}$) & (3 - 9,32;~1,16;~ 5,1)   \\ \hline
		Optimized DN-CNN (Hidden layers -  $n_{l},f_{l}$) & (7 - 16,~3)   \\
		\hline
		SR-ConvLSTM (Hidden layers -  $n_{l},f_{l}$) & (3 - 9,64;~1,32;~ 5,1)   \\
		\hline
		Activation function              & ReLU                      \\ \hline
		Number of epochs        & 250                                \\ \hline
		Training samples        & 8000                             \\ \hline
		Testing samples        & 2000                             \\ \hline
		Batch size          & 128                                    \\ \hline
		Optimizer       & ADAM                                       \\ \hline
		Loss function      & MSE                                     \\ \hline
		Learning rate        & 0.001                                 \\ \hline
		Training SNR        & 40 dB                                 \\ \hline
	\end{tabular}
\end{table} 

\subsection{TS-ChannelNet}
\ac{TS-ChannelNet}~\cite{ref_TS_ChannelNet}  is based on applying {\ac{ADD-TT}} interpolation to the received {\ac{OFDM}} frame. Thereafter, accurate estimation is achieved by implementing {\ac{SR-ConvLSTM}} network to track doubly-dispersive channel variations by learning the vehicular channel's time and frequency correlations.
It is observed that the {\ac{ADD-TT}} interpolation is an {\ac{SBS}} estimator, where {\ac{DPA}} estimation is initially applied as explained in~\eqref{eq: DPA_1} and~\eqref{eq: DPA_2}. Thereafter, the enlarged  {\ac{DPA}} demapping error is reduced by applying time domain truncation in the following manner
\begin{equation}
\hat{{\ma{h}}}_{\text{DPA}_{i}} = {\ma{F}}_{\text{K}}^{\text{H}} \hat{\tilde{\ma{h}}}_{\text{DPA}_{i}},
\label{eq:IFFT}
\end{equation}
where ${\ma{F}}_K \in \compl^{K \times K}$ denotes the $K$-DFT matrix, and $\hat{{\ma{h}}}_{\text{DPA}_{i}}$ represents the time-domain {\ac{DPA}} estimated channel. Thereafter, $\hat{{\ma{h}}}_{\text{DPA}_{i}}$ truncation is applied to the significant $L$ channel taps, such that  
\begin{equation}
	\hat{{\ma{h}}}_{\text{DPA}_{i,L}} =  \hat{{\ma{h}}}_{\text{DPA}_{i}} ( 1:L).
	\label{eq:IFFT_TT}
\end{equation}
Next, $\hat{{\ma{h}}}_{\text{DPA}_{i,L}}$ is converted back to the frequency domain such that
\begin{equation}
\hat{\tilde{\ma{h}}}_{\text{TT}_{i}} = {\ma{F}}_{\text{K}} \hat{{\ma{h}}}_{\text{DPA}_{i,L}},
\label{eq:TT}
\end{equation}
%

Implementing the average time truncation operation to $\hat{\tilde{\ma{h}}}_{\text{DPA}_{i}}[k]$ lowers the effect of noise and enlarged demapping error. Moreover, 
$\hat{\tilde{\ma{h}}}_{\text{TT}_{i}}[k]$ estimated channel is further enhanced by applying frequency and time-domain averaging consecutively as follows
\begin{equation}
\hat{\tilde{\ma{h}}}_{\text{FTT}_{i}}[k] = \sum_{\lambda = -\beta}^{\lambda = \beta} \omega_{\lambda} \hat{\tilde{\ma{h}}}_{\text{TT}_{i}}[k + \lambda], ~ \omega_{\lambda} = \frac{1}{2\beta+1}.
\label{eq: STA_44}
\end{equation}
The final {\ac{ADD-TT}} channel estimates are updated using time averaging between the previously {\ac{ADD-TT}} estimated channel and the frequency averaged channel in~{\eqref{eq: STA_44}}, such that
\begin{equation}
\hat{\tilde{\ma{h}}}_{\text{ADD-TT}_{i}}[k] = (1 - {\alpha})  \hat{\tilde{\ma{h}}}_{\text{ADD-TT}_{i-1}}[k] + {\alpha}\hat{\tilde{\ma{h}}}_{\text{FTT}_{i}}[k].
\label{eq: STA_55}
\end{equation}
The doubly-dispersive channel can be modeled as a time-series forecasting problem. Here, historical data can be utilized to forecast future observations~\cite{ref_timeseriers}. Motiviated by this possibility, the authors in~\cite{ref_TS_ChannelNet} apply \ac{SR-ConvLSTM} network in addition to the {\ac{ADD-TT}} interpolation, where convolutional layers get added to the LSTM network to capture more doubly-dispersive channel features. Consequently, this improves the  estimation performance. Accordingly, the {\ac{ADD-TT}} estimated channel for the entire received frame is modeled as a low resolution image. Next, the \ac{SR-ConvLSTM} network is used after the {\ac{ADD-TT}} interpolation. Unlike {\ac{ChannelNet}} estimator where two {\acp{CNN}} are employed, {\ac{TS-ChannelNet}} estimator uses only one {\ac{SR-ConvLSTM}} network, which relatively reduces the overall computational complexity. However, {\ac{TS-ChannelNet}} continues to be ridden with high computational complexity due to the integration of LSTM and {\ac{CNN}} in a single network.

\begin{figure*}[t]
	\centering
	\includegraphics[height=8cm,width=2\columnwidth]{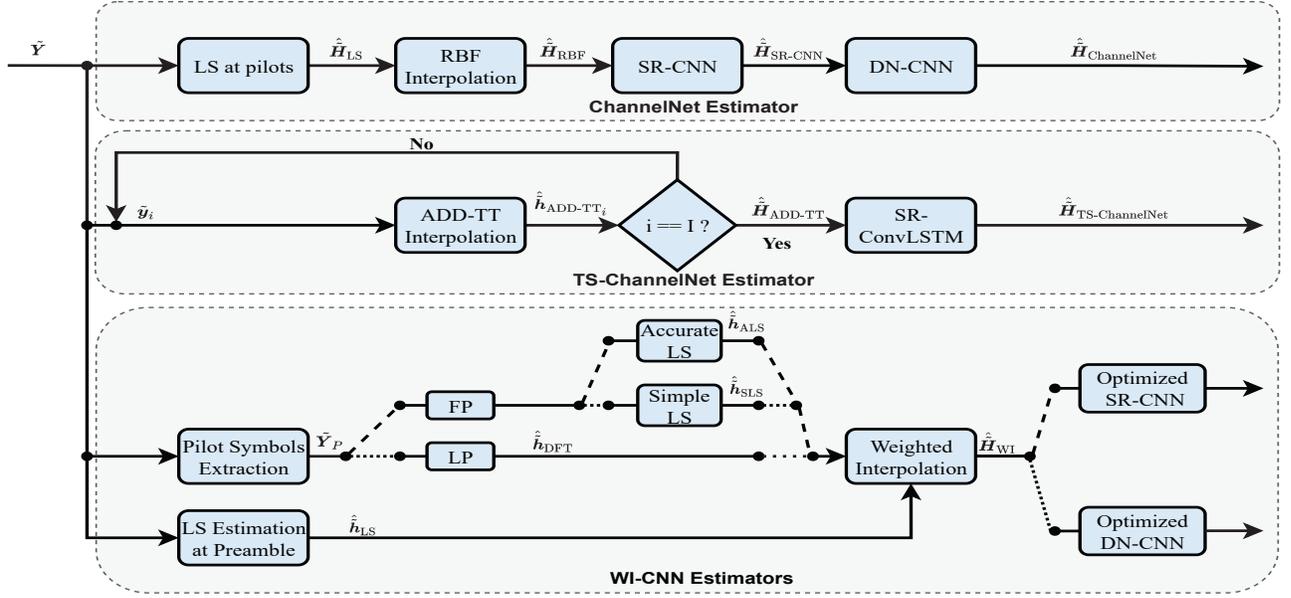}
	\caption{The block diagram of the studied CNN-based FBF channel estimators.}
	\label{fig:soa_Figure}
\end{figure*}

\subsection{WI-CNN}

To overcome the limitations of the {\ac{ChannelNet}} and \ac{TS-ChannelNet} estimators, \ac{WI}-{\ac{CNN}} estimator has been proposed in~\cite{ref_WI_CNN}. In this method,  the frame structure is adapted in accordance with the mobility condition employing varied pilot allocation schemes. Particularly, only $P$ pilot {\ac{OFDM}} symbols are required in the transmitted frame, such that $\tilde{\ma{Y}}_{P} = [\tilde{\ma{y}}^{(p)}_{1}, \dots,  \tilde{\ma{y}}^{(p)}_{q}, \dots, \tilde{\ma{y}}^{(p)}_{P}] \in \compl ^{K_{\text{on}}\times P}$. The index $1 \leq q \leq P$ refers to the location of the \ac{OFDM} pilot symbol in the frame. The other $I_{d} = I - P$ {\ac{OFDM}} data symbols are employed for  data transmission purposes. As per the employed pilots allocation scheme, the channel is estimated at the inserted pilot symbols, after which {\ac{WI}} is applied to estimate the channel at the {\ac{OFDM}} data symbols. The estimated frame is then modeled as a 2D noisy image where optimized {\ac{SR-CNN}} and {\ac{DN-CNN}} are utilized for noise elimination. Against this backdrop, the {\ac{WI}}-{\ac{CNN}} proceeds as follows
\begin{itemize}
    \item  Channel estimation at the pilot symbols:  Two pilot allocation schemes are defined. The full pilot allocation ($\text{FP}$) where $K$ pilots are inserted within all pilot symbols and {\ac{LS}} estimation is applied to estimate the channel for each inserted pilot symbol, such that
        \begin{equation}
        \hat{\tilde{\ma{h}}}_{{\text{SLS}}_{q}}[k] = \frac{\tilde{\ma{y}}^{(p)}_{q}[k]}{\tilde{\ma{p}}[k]}.
        \label{eq: SLSP}
        \end{equation}
        $\hat{\tilde{\ma{h}}}_{{\text{SLS}}_{q}}[k]$ represents the  {\ac{SLS}} estimation at the $q$-th inserted pilot symbol. In addition, the {{\ac{ALS}}} that can be obtained by implementing the {\ac{DFT}} interpolation of estimated channel impulse response at the $q$-th received pilot symbol $\hat{\ma{h}}_{q,L}$, such that
        \begin{equation}
        \hat{\tilde{\ma{h}}}_{{\text{ALS}}_{q}} = \ma{F}_{\text{K}} \hat{\ma{h}}_{q,L},~~ \hat{\ma{h}}_{q,L} = \ma{F}_{\text{K}}^{\dagger} \hat{\tilde{\ma{h}}}_{{\text{LS}}_{q}}.
        \label{eq:ALSP}
        \end{equation}
        {\ac{ALS}} relies on the fact that $\tilde{\ma{h}}_{{q}} =\ma{F}_{\text{K}} {\ma{h}}_{q,L}$, where ${\ma{h}}_{q,L} \in \compl ^{L\times 1}$ signifies the channel impulse response at the $q$-th received pilot symbol that can be estimated by employing the pseudo inverse matrix of $\ma{F}_{\text{K}}$, namely,  $\ma{F}_{\text{K}}^{\dagger} = [(\ma{F}_{\text{K}}^{\text{H}} \ma{F}_{\text{K}})^{-1} \ma{F}_{\text{K}}^{\text{H}}]$ .
        However, if the number of doubly dispersive-channel taps $L$ remains known, only $K_{p} = L$ pilot subcarriers are sufficient in each inserted pilot symbol. Accordingly,~\eqref{eq:ALSP} can be rewritten as
        \begin{equation}
        \hat{\tilde{\ma{h}}}_{{\text{DFT}}_{q}} = \ma{F}_{\text{K}} \hat{\ma{h}}_{q,L},~~ \hat{\ma{h}}_{q,L} = \ma{F}_{p}^{\dagger} \hat{\tilde{\ma{h}}}_{{\text{LS}}_{q}}.
        \label{eq:ALSPL}
        \end{equation}
         
         $\ma{F}_{p}^{\dagger} = [(\ma{F}_{p}^{\text{H}} \ma{F}_{p})^{-1} \ma{F}_{p}^{\text{H}}]$ denotes the pseudo inverse matrix of $\ma{F}_{p} \in \compl^{K_{\text{p}} \times L}$ referring to the truncated DFT matrix obtained by selecting $\Kp$ rows, and $L$ columns from the $K$-DFT matrix.

    \item Channel estimation at data symbols: The estimated channels of the $P$ pilot symbols are first grouped into $P$ matrices to estimate the channel for each received OFDM data symbol, such that
      
      \begin{equation}
          \hat{\tilde{\ma{H}}}_{q} = [\hat{\tilde{\ma{h}}}_{q-1}, \hat{\tilde{\ma{h}}}_{q}],~ q = 1, \cdots P.
      \end{equation}
      
      $\hat{\tilde{\ma{h}}}_{0} = \hat{\tilde{\ma{h}}}_{\text{LS}}$ refers to the \ac{LS} estimated channel at the beginning of the received frame~\eqref{eq: LS}. Thus, the received frame can be divided into $P$ sub-frames, where $f$ refers to the sub-frame index, such that $1 \leq f \leq P$. Therefore, the estimated channel for the $i$-th received {\ac{OFDM}} symbol within each $f$-th sub-frame can be expressed as follows
      \begin{equation}
          \hat{\tilde{\ma{H}}}_{{\text{WI}}_{f}} =  \hat{\tilde{\ma{H}}}_{{f}} \ma{C}_{f}.
      \label{eq:WI_LS}
      \end{equation}

      $\hat{\tilde{\ma{H}}}_{f} \in \compl^{ K \times 2} $ denotes the  {\ac{LS}} estimated channels at the pilot symbols within the $f$-th sub-frame, and $\ma{C}_{f} \in \mathbb{R}^{2 \times I_{f}}$ the interpolation weights of the $I_{f}$ {\ac{OFDM}} data symbols within the $f$-th sub-frame. The interpolation weights $\ma{C}_{f}$ are calculated by minimizing the {\ac{MSE}} between the ideal channel $\tilde{\ma{H}}_{{f}}$, and the {\ac{LS}} estimated channel at the {\ac{OFDM}} pilot symbols $\hat{\tilde{\ma{H}}}_{{f}}$ as obtained in~\cite{ref_interpolation_matrix} and expressed in~\eqref{eq:ci}. There, $J_{0}(.)$ is the zeroth order Bessel function of the first kind, $T_{\text{s}}$ signifies the received {\ac{OFDM}} data symbol duration, whereas $E_{{{q}}}$ denotes the overall noise of the estimated channel at the $q$-th pilot symbol.
      
    \begin{figure*}[h]
    \begin{equation}
    \begin{split}
     \ma{C}_{{f}} &= \Ex{\tilde{\ma{H}}_{{f}} \hat{\tilde{\ma{H}}}^{H}_{{f}}} \left[\Ex{\hat{\tilde{\ma{H}}}_{{f}} \hat{\tilde{\ma{H}}}^{H}_{{f}}} \right]^{-1} = \begin{bmatrix} \Ex{ \tilde{\ma{H}}_{{f}} \hat{\tilde{\ma{h}}}^{H}_{{q}}} & \Ex{ \tilde{\ma{H}}_{{i}} \hat{\tilde{\ma{h}}}^{H}_{{q+1}}} \end{bmatrix} \begin{bmatrix}
     \Ex{\norm{{\tilde{\ma{h}}}_{{q}}}^2} + E_{{{q}}} & \Ex{{\tilde{\ma{h}}}_{{q}} {\tilde{\ma{h}}}^{H}_{{q+1}}}\\
     \Ex{{\tilde{\ma{h}}}_{{q+1}} {\tilde{\ma{h}}}^{H}_{{q}}} & \Ex{\norm{{\tilde{\ma{h}}}_{{q+1}}}^2} + E_{{{q+1}}}
     \end{bmatrix}^{-1} \\
     &=    \begin{bmatrix}
     J_{0} (2 \pi f_{\text{d}} (f-1) T_{\text{s}}) &
     J_{0} (2 \pi f_{\text{d}} (I_{f} + 1 - f) T_{\text{s}})
     \end{bmatrix}  \begin{bmatrix}
     1 + E_{{{\Phi}_{q}}} & J_{0} (2 \pi f_{\text{d}} I_{f} T_{\text{s}})\\
     J_{0} (2 \pi f_{\text{d}} I_{f} T_{\text{s}}) & 1 + E_{{{q+1}}}
     \end{bmatrix}^{-1}.
    \end{split}
    \label{eq:ci}
    \end{equation}
    \end{figure*}
    \item CNN-based Processing: The final step in the {\ac{WI}}-{\ac{CNN}} estimators is to apply {\ac{CNN}} processing to further improve the {\ac{WI}} estimated channels. Optimized {\ac{SR-CNN}} and {\ac{DN-CNN}} are employed in this context. The investigations conducted in~\cite{ref_WI_CNN} reveal that both {\ac{SR-CNN}} and {\ac{DN-CNN}} networks have similar performance in low mobility scenarios, whereas {\ac{DN-CNN}} outperforms {\ac{SR-CNN}} in high mobility scenarios. \figref{fig:soa_Figure} and Table~\ref{tb:CNN_params} illustrate the block diagram as well as configured parameters of the studied CNN-based channel estimators, respectively. Furthermore, the salient features of the studied DL-based channel estimators are summarized in Table~\ref{tb:SoA_Summary}. Notably, robustness feature alludes to the ability of the studied estimation to maintain good performance as the variation of the doubly-dispersive channel increases.
\end{itemize}

\begin{table*}[t]
\renewcommand{\arraystretch}{1.3}
\small
\caption{The characteristics of the employed vehicular channel models  following Jake's Doppler spectrum.}
\label{tb:VCMC}
\begin{tabular}{|c|c|c|c|c|c|}
\hline
\begin{tabular}{@{}c@{}}\textbf{Channel} \\\textbf{model} \end{tabular}  & \begin{tabular}{@{}c@{}}\textbf{Channel} \\\textbf{taps} \end{tabular} & \begin{tabular}{@{}c@{}}\textbf{Vehicle velocity} \\\textbf{[kmph]} \end{tabular} & \begin{tabular}{@{}c@{}}\textbf{Doppler} \\\textbf{shift [Hz]} \end{tabular}  & \textbf{Average path gains {[}dB{]}}                                                                                       & \textbf{Path delays {[}ns{]}}                                                                            \\ \hline
VTV-UC                 & 12                             & 45                                & 250                                 & \begin{tabular}[c]{@{}c@{}}{[}0, 0, -10, -10, -10, -17.8, -17.8,\\ -17.8, -21.1, -21.1, -26.3, -26.3{]}\end{tabular}     & \begin{tabular}[c]{@{}c@{}}{[}0, 1, 100, 101, 102, 200, 201,\\202, 300, 301, 400, 401{]}\end{tabular} \\ \hline
VTV-SDWW               & 12                             & 100-200                                  & 500-1000                                & \begin{tabular}[c]{@{}c@{}}{[}0, 0, -11.2, -11.2, -19, -21.9, -25.3,\\ -25.3, -24.4, -28, -26.1, -26.1{]}\end{tabular} & \begin{tabular}[c]{@{}c@{}}{[}0, 1, 100, 101, 200, 300, 400,\\401, 500, 600, 700, 701{]}\end{tabular} \\ \hline
\end{tabular}
\end{table*}


\section{Simulation Results} \label{simulation_results}
This section illustrates the performance evaluation of the studied {\ac{DL}}-based SBS and FBF estimators in relation to \ac{BER}, \ac{NMSE} employing varied metrics and mobility scenarios. 
%
%
\subsection{Configuration Setup}
To simulate doubly-dispersive channels, vehicular communications is considered a simulation case study, where three \ac{TDL} channel models~\cite{ref_channel1} are defined as follows
\begin{itemize}
    \item Low mobility: where VTV Urban Canyon (VTV-UC) vehicular channel model is considered. This channel model is measured between two vehicles moving in a dense urban traffic environment at ${V} = 45$ Kmph equivalent to  ${f}_{d} = 250$ Hz.
    
    \item High and very high mobility: These scenarios measure the communication channel between two vehicles moving on a highway having center wall between its lanes at ${V} = 100$ Kmph and $200$ Kmph equivalent to ${f}_{d} = 500$ Hz and ${f}_{d} = 1000$ Hz, respectively. This vehicular channel model is referred to as VTV Expressway Same Direction with Wall (VTV-SDWW).     
\end{itemize}

The employed channel models are generated after the wide-sense stationary uncorrelated scattering (WSSUS) model~{\cite{ref1400}}. Thus, we have
\begin{itemize}
    \item  Each path $h_{l}(t)$ is a zero mean Gaussian complex process, $E\{h_{l}(t)\} = 0, \forall t$, and the mean of each path is independent of the time variations. Moreover, the time correlation function $r_{h_{l}}(t_{1}, t_{2}) = E\{h_{l}(t_{1})h^{*}_{l}(t_{2})\}$ can only be written with the difference $\Delta(t) = (t_{1} - t_{2})$, such that
    
    \begin{equation}
        r_{h_{l}}(t_{1}, t_{2}) =  r_{h_{l}}(\Delta_t).
        \label{eq:tc}
    \end{equation}
    
    Then, each path $h_{l}(t)$ is the wide sense stationary (WSS).
    
    \item Uncorrelated scattering (US) implies that the paths are uncorrelated, so for $l_{1} \ne l_{2}$ we have
    \begin{equation}
        \small
        E[h_{l_{1}}(t)h^{*}_{l_{2}}(t)] = 0.
        \label{eq:US}
    \end{equation}
\end{itemize}

Table~{\ref{tb:VCMC}} illustrates the main characteristics of the defined {\ac{TDL}} channel models.

The {\ac{OFDM}} simulation parameters are based on the IEEE 802.11p standard as illustrated in~Table~\ref{tb:IEEE80211p_specs}. These simulations are implemented using QPSK and 16QAM modulation orders, the SNR range is $[0 ,5 ,\dots, 40 ]$ dB. In addition, the performance evaluation is made according to: (\textit{i}) modulation order, (\textit{ii}) mobility, (\textit{iii}) frame length, and (\textit{iv}) DL architecture.

Finally, it is observed that the conventional 2D {\ac{LMMSE}} estimator~\cite{ref_lmmse} is included in the performance evaluation of the DL-based FBF estimators as a lower bound performance limit. The 2D LMMSE estimator almost achieves a similar performance as the ideal channel, but is ridden with high computational complexity. This renders it impractical in terms of real-time applications.

\begin{table}[t]
	\renewcommand{\arraystretch}{1.3}
	\small
	\centering
	\caption{Simulation parameters of the IEEE 802.11p physical layer.}
	\label{tb:IEEE80211p_specs}
	\begin{tabular}{|c|c|}
		\hline
		\textbf{Parameter}      & \textbf{IEEE 802.11p}                    \\ \hline
		Bandwidth               & 10 MHz  \\ \hline
		Guard interval duration & 1.6~$\mu\mbox{s}$       \\ \hline
		Symbol duration         & 8~$\mu\mbox{s}$                 \\ \hline
		Short training symbol duration         & 1.6~$\mu\mbox{s}$    \\ \hline
		Long training symbol duration         & 6.4~$\mu\mbox{s}$ \\ \hline
		Total subcarriers      & 64     \\ \hline
		Pilot subcarriers      & 4  \\ \hline
		Data subcarriers       & 48     \\ \hline
		Subcarrier spacing     & 156.25 KHz       \\ \hline
	\end{tabular}
\end{table}

\subsection{DL-Based SBS Estimation Schemes}

\subsubsection{Modulation Order}

\begin{figure*}[t]
	\setlength{\abovecaptionskip}{3pt plus 3pt minus 2pt}
	\centering
	\includegraphics[width=2\columnwidth]{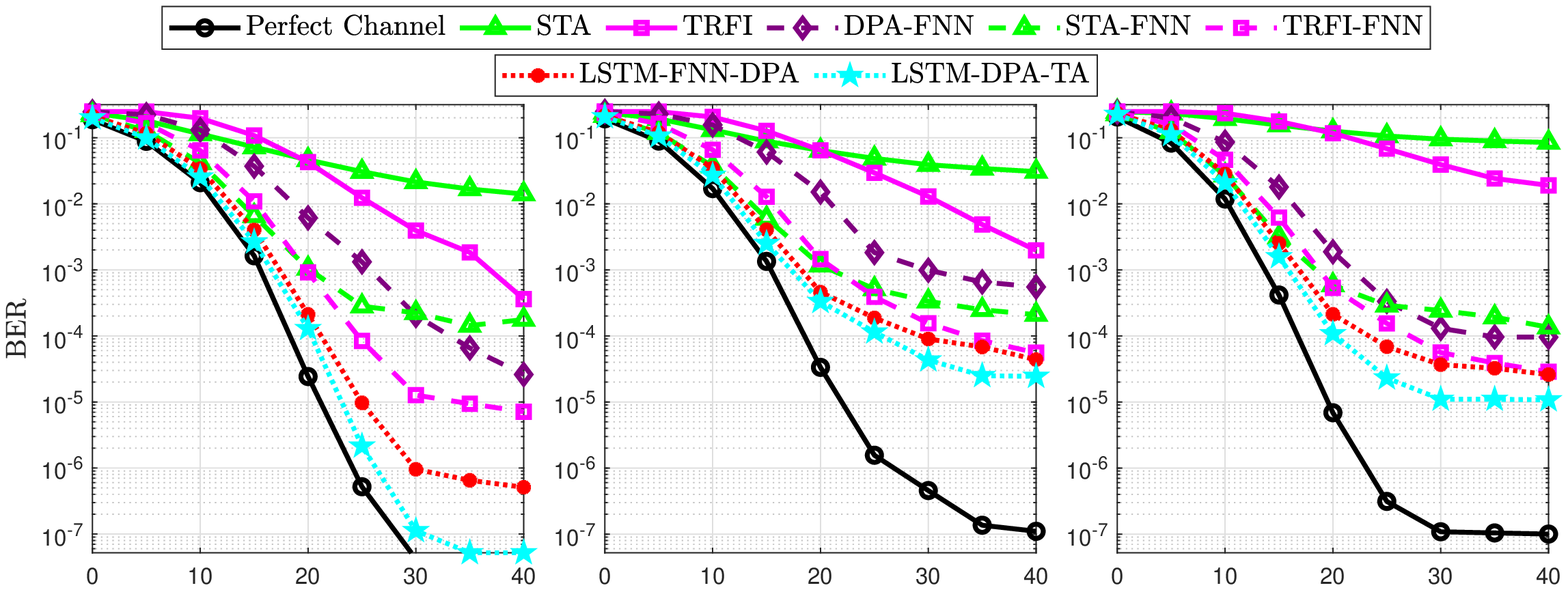}\\[-3ex]
	\subfloat[\label{BER_QPSK_DL_SBS} BER performance employing QPSK.]{\hspace{.5\linewidth}} \\[-2ex]
	\vspace*{10pt}
	\includegraphics[width=2\columnwidth]{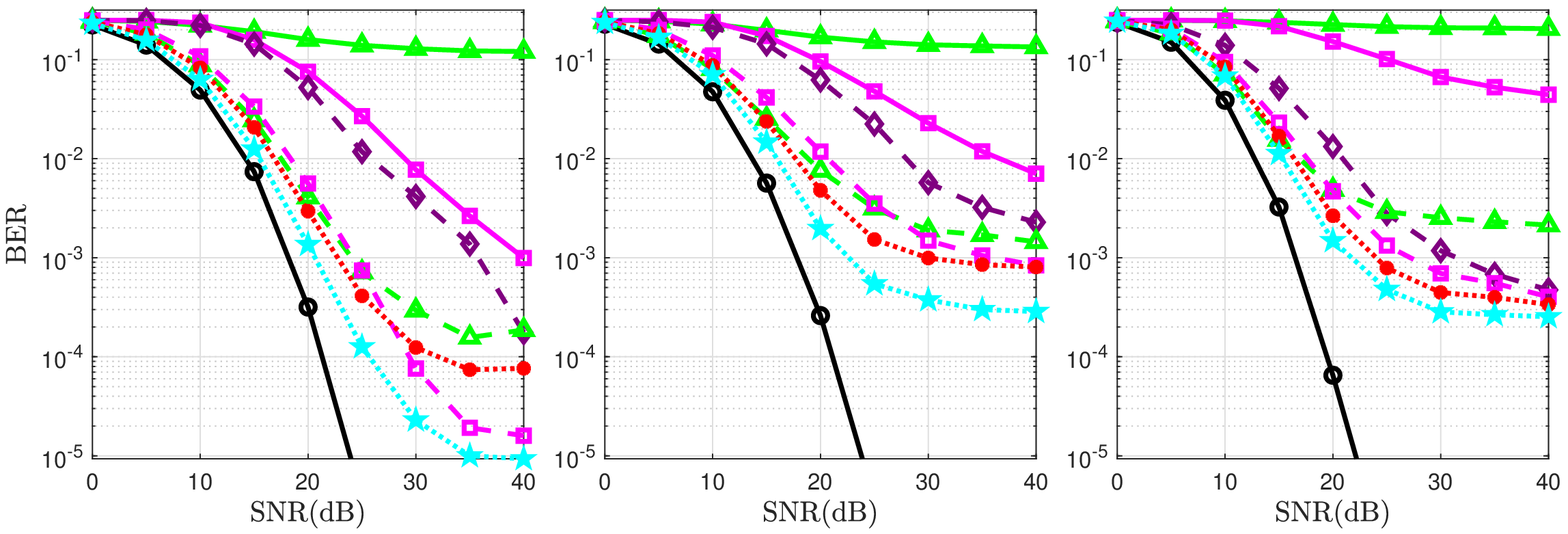}\\[-3ex]
	\subfloat[\label{BER_16QAM_DL_SBS} BER performance employing 16QAM.]{\hspace{.5\linewidth}}
	\caption{BER for $I=100$, mobility from left to right: low  ($v = 45~\text{Kmph}, f_{d} = 250$ Hz), high  ($v = 100~\text{Kmph}, f_{d} = 500$ Hz), very high  ($v = 200~\text{Kmph}, f_{d} = 1000$ Hz).}
	\label{fig:BER_DL_SBS}
\end{figure*}

\begin{figure*}[t]
	\setlength{\abovecaptionskip}{3pt plus 3pt minus 2pt}
	\centering
	\includegraphics[width=2\columnwidth]{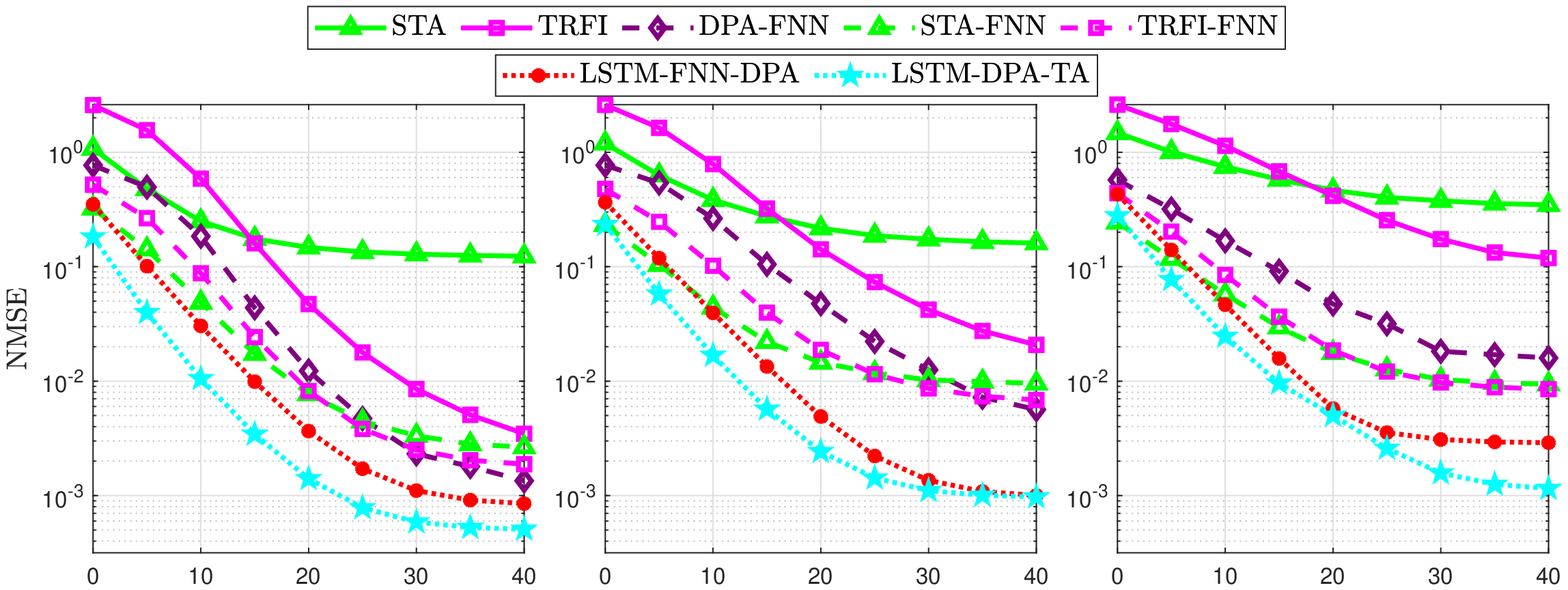} 
 	\vspace*{10pt}
	\caption{NMSE for $I=100$, mobility from left to right: low  ($v = 45~\text{Kmph}, f_{d} = 250$ Hz), high  ($v = 100~\text{Kmph}, f_{d} = 500$ Hz), very high  ($v = 200~\text{Kmph}, f_{d} = 1000$ Hz).}
	\label{fig:NMSE_DL_SBS}
\end{figure*}

For QPSK modulation order, we can notice from \figref{fig:BER_DL_SBS}, and {\figref{fig:NMSE_DL_SBS} that conventional {\ac{SBS}} estimators witness a considerable performance degradation in different mobility scenarios primarily due to the enlarged DPA demapping error, particularly under very high mobility. Nevertheless, employing DL techniques in the channel estimation process results in a significant improvement in overall performance. To begin with, the FNN-based estimators, where FNN is employed as a post-processing unit after conventional estimators, are discussed. As observed, FNN can implicitly learn the channel correlations apart from preventing a high demapping error arising from conventional DPA-based estimation, while STA-FNN and TRFI-FNN outperform conventional STA and TRFI estimators by at least $15$ dB gain in terms of SNR for BER $= 10^{-3}$. Meanwhile, STA-FNN estimator outperforms DPA-FNN estimator by around $5$ dB gain in terms of SNR for BER $= 10^{-3}$. However, STA-FNN suffers from error floor beginning from SNR $= 20$ dB, particularly in very high mobility scenarios. This is attributed to the fact that conventional STA estimation outperforms DPA in low SNR region due to the frequency and time averaging operations that can alleviate the impact of noise and demapping error in low SNR regions. On the other hand, the averaging operations are not useful in high SNR regions since the impact of noise is low, and the STA averaging coefficients are fixed. Therefore, TRFI-FNN is used to improve the performance at high SNRs to compensate for the STA-FNN performance degradation in high SNR region. Importantly, STA-FNN and TRFI-FNN can be employed in an adaptive manner where STA-FNN and TRFI-FNN are used in low and high SNR regions, respectively. 

For the LSTM-based estimators, employing LSTM as a prepossessing unit rather than a simple FNN in the channel estimation has shown to bring about a significant improvement in the overall performance. This is because LSTM is capable of efficiently learning the time correlations of the channel by taking the advantage of the previous output apart from the current input in order to estimate the current output. LSTM-FNN-DPA estimator~\cite{ref_lstm_dnn_dpa} outperforms  STA-FNN and TRFI-FNN estimators by approximately $4$ dB gain in terms of SNR for BER $= 10^{-3}$. However, this estimator is not impervious to high computational complexity, as discussed in the next section, due to the utilization of two DL networks, i.e, LSTM followed by FNN. {On the other hand, the LSTM-DPA-TA estimators performance gain in various scenarios can be explained by employing the {\ac{TA}} processing, which significantly alleviates the noise impact aside from the strong ability of the LSTM in learning the channel time correlations compared with a simple FNN architecture.} The LSTM-DPA-TA estimator outperforms the LSTM-FNN-DPA estimator by around $4$ dB gain in terms of SNR for BER $= 10^{-4}$. When adopting high modulation order (16QAM), the LSTM-DPA-TA estimator outperforms the other estimators by at least $7$ dB and $3$ dB gains in terms of SNR for BER $= 10^{-3}$ in high as well as very high mobility scenarios, respectively, as illustrated in \figref{BER_16QAM_DL_SBS}.

\subsubsection{Mobility}
The degraded performance with the increased mobility of all the studied schemes can be observed from \figref{fig:BER_DL_SBS}. However, the time diversity gain increases when there is an increase in the Doppler spread, as evidenced by comparing the case of the DL-based estimators in high mobility $(f_d = 500)$ and very high mobility ($f_d = 1000$). This behavior can be explained by the ability of DL networks to reduce the estimation error stemming from the AWGN noise and the DPA demapping error. By contrast, the net gain from the time diversity is influenced by the AWGN noise and DPA demapping error, as is the case in conventional SBS estimators. The performance degradation is attributed as the mobility increases since the impact of the AWGN noise and DPA demapping error is much more dominant than the time diversity gain. This observation is also valid for high modulation orders such as 16QAM, as evidenced in \figref{BER_16QAM_DL_SBS}.
\begin{figure*}[t]
	\setlength{\abovecaptionskip}{3pt plus 3pt minus 2pt}
	\centering
	\includegraphics[width=2\columnwidth]{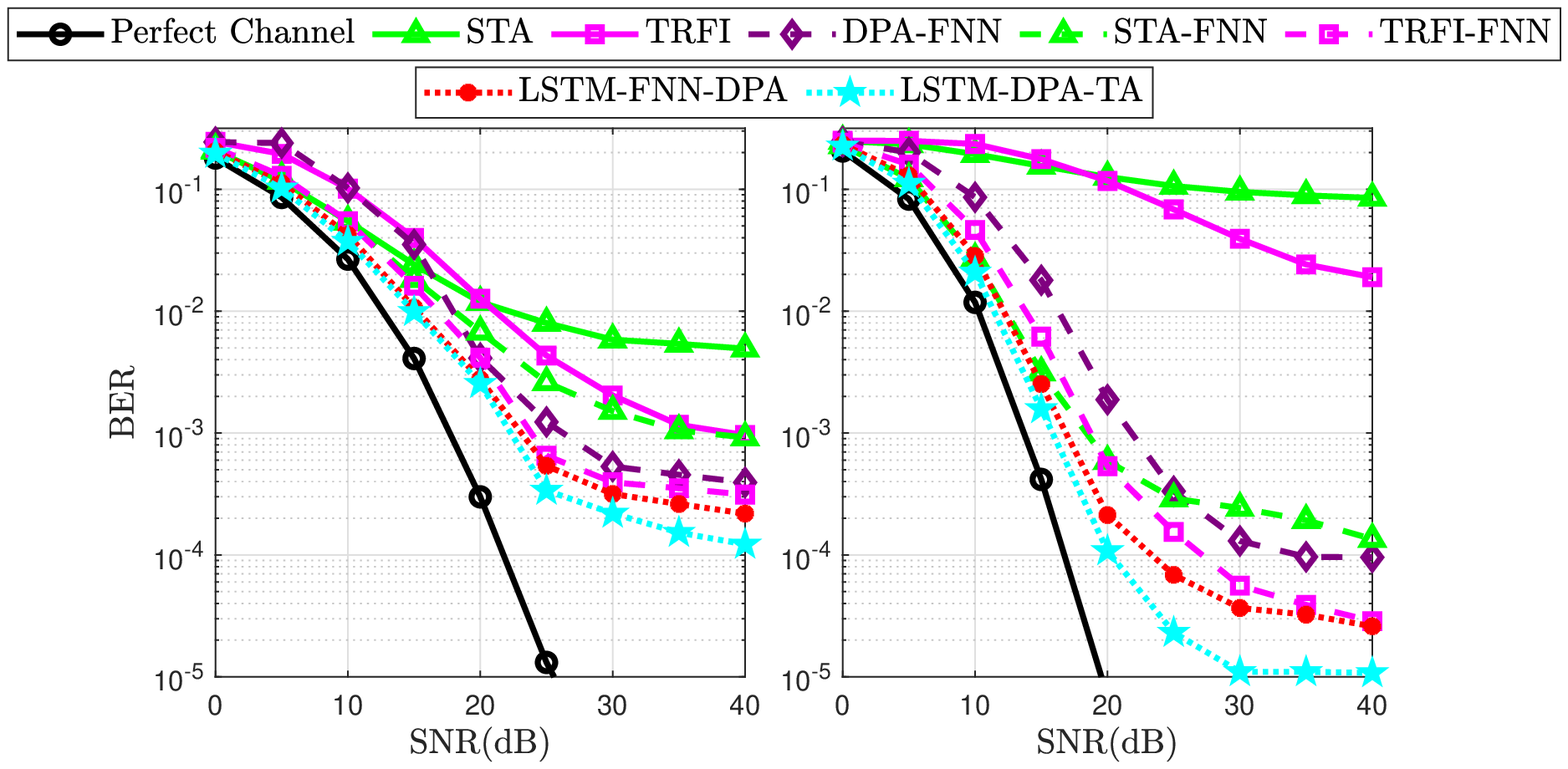} 
	\vspace*{10pt}
	\caption{BER for QPSK, very high mobility  ($v = 200~\text{Kmph}, f_{d} = 1000$ Hz) from left to right: $I = 10$, $I = 100$.}
	\label{fig:BER_QPSK_DL_SBS_FL}
\end{figure*}
\begin{figure*}[t]
	\setlength{\abovecaptionskip}{3pt plus 3pt minus 2pt}
	\centering
	\includegraphics[width=2\columnwidth]{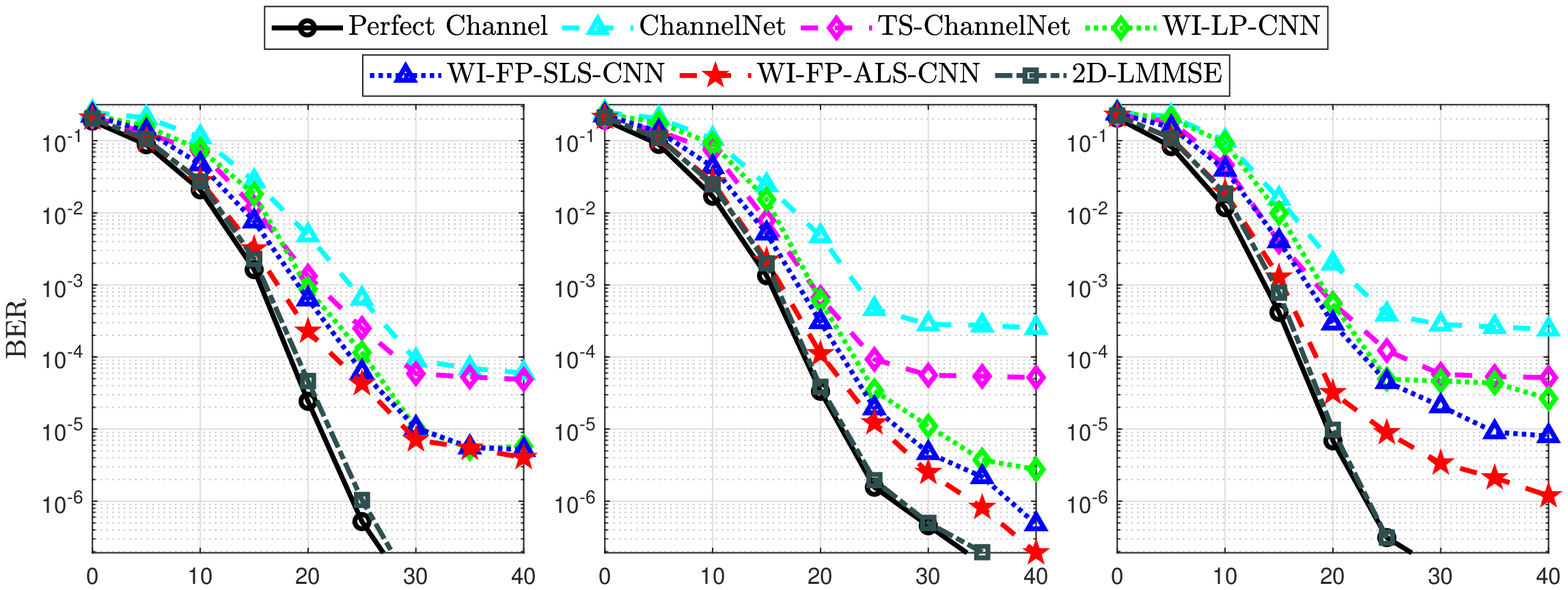}\\[-3ex]
	\subfloat[\label{BER_QPSK_DL_FBF} BER performance employing QPSK.]{\hspace{.5\linewidth}} \\[-2ex]
	\vspace*{10pt}
	\includegraphics[width=2\columnwidth]{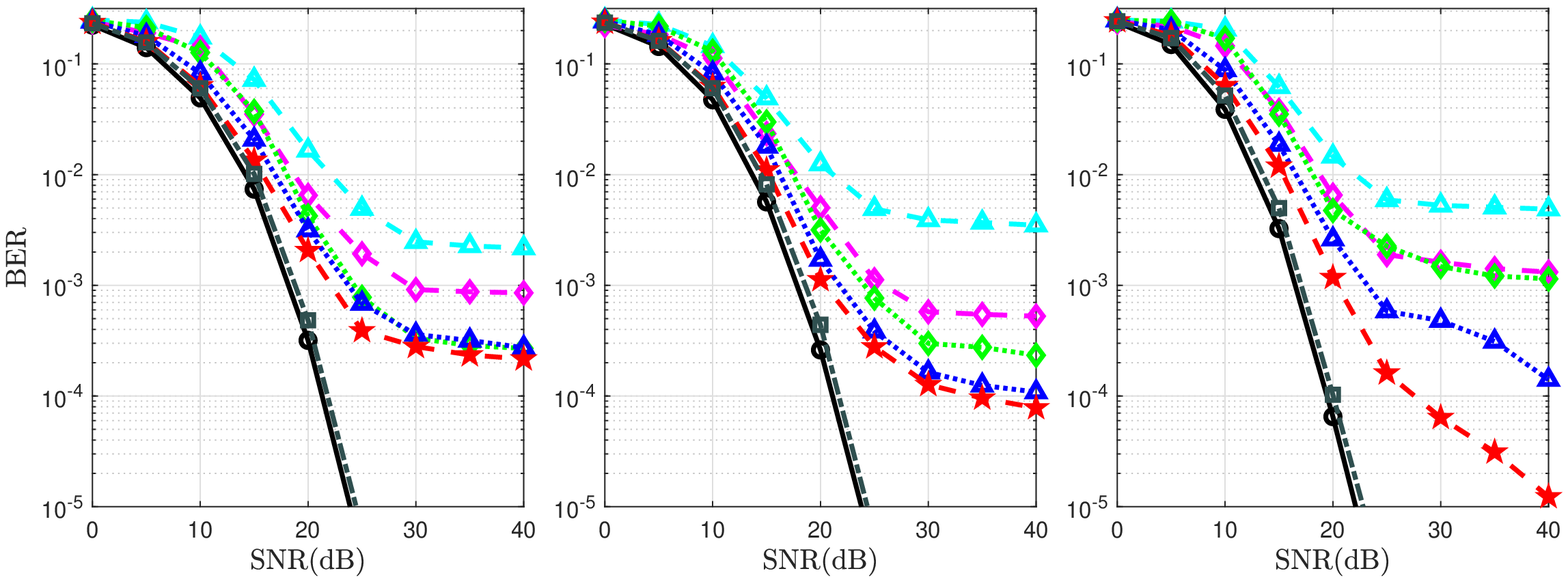}\\[-3ex]
	\subfloat[\label{BER_16QAM_DL_FBF} BER performance employing 16QAM.]{\hspace{.5\linewidth}}
	\caption{BER for $I=100$, mobility from left to right: low  ($v = 45~\text{Kmph}, f_{d} = 250$ Hz), high  ($v = 100~\text{Kmph}, f_{d} = 500$ Hz), very high  ($v = 200~\text{Kmph}, f_{d} = 1000$ Hz). The CNN refers to SR-CNN and DN-CNN in low and high/very high) mobility scenarios, respectively.}
	\label{fig:BER_DL_FBF}
\end{figure*}
\subsubsection{Frame Length\label{Results_Mobility_DL}}
The impact of frame length is illustrated in \figref{fig:BER_QPSK_DL_SBS_FL}. As can be seen, the performance of the conventional estimators strongly depends on the frame length, given that employing short frame $I = 10$ results in a negligible accumulated DPA demapping error. By contrast, the {DL}-based estimators are found to be more robust against the changes in the employed frame length. However, in the case of a long frame ($I = 100$), the performance gain of the DL-based estimators is significantly remarkable. This behavior is mainly attributed to the time diversity negligible gain when short frame is employed and vice versa.   

To conclude, it can be surmised that increasing the frame length increases the time diversity gain. Additionally, the codeword becomes longer with a longer frame. Therefore, the time diversity is capable of compensating for Doppler error, particularly in very high mobility scenario as illustrated in \figref{fig:BER_QPSK_DL_SBS_FL}


\subsubsection{DL Architecture \label{Results_DL_Architecture_DL}}
The DPA-FNN estimator integrates three hidden layer FNN in additon to the conventional DPA estimation with $40-20-40$ neurons. However, as can be observed in \figref{fig:BER_DL_SBS}, correcting the estimation error of the DPA estimation is insufficient even after the inclusion of more neurons in the FNN hidden layers, because it merely corrects the demapping error, neglecting the received symbols'requency and time correlation. 
Meanwhile, the {\ac{STA}}-{\ac{FNN}} and TRFI-FNN estimators have better optimized three hidden layers {\ac{FNN}} architecture where $15-15-15$ neurons are used. Consequently, the overall computational complexity is considerably lowered when compared to the DPA-FNN, while attaining performance superiority. This is due to the fact that \ac{STA} considers frequency as well a time correlation between the received \ac{OFDM} symbols, while the conventional TRFI estimator employs frequency-domain cubic interpolation to make further improvements in the DPA estimation.

Therefore, it can be concluded that the pre-estimation should be good enough in order for the FNN processing to be more useful. Put differently, with an increased accuracy of the pre-estimation, low-complexity FNN architecture can be taken advantage of while recording a significant performance gain. On the contrary, if the pre-estimation is poor, employing FNN processing with high-complexity architecture results in a limited performance gain while increasing the overall computational complexity. As is the case with LSTM-based estimators, employing the {\ac{TA}} processing in the LSTM-DPA-TA estimator to ameliorate the AWGN noise impact results in a less complex architecture in comparison to the LSTM-FNN-DPA estimator, where two {\ac{DL}} networks are employed.
\subsection{DL-Based FBF estimation Scheme}
\begin{figure*}[t]
	\setlength{\abovecaptionskip}{3pt plus 3pt minus 2pt}
	\centering
	\includegraphics[width=2\columnwidth]{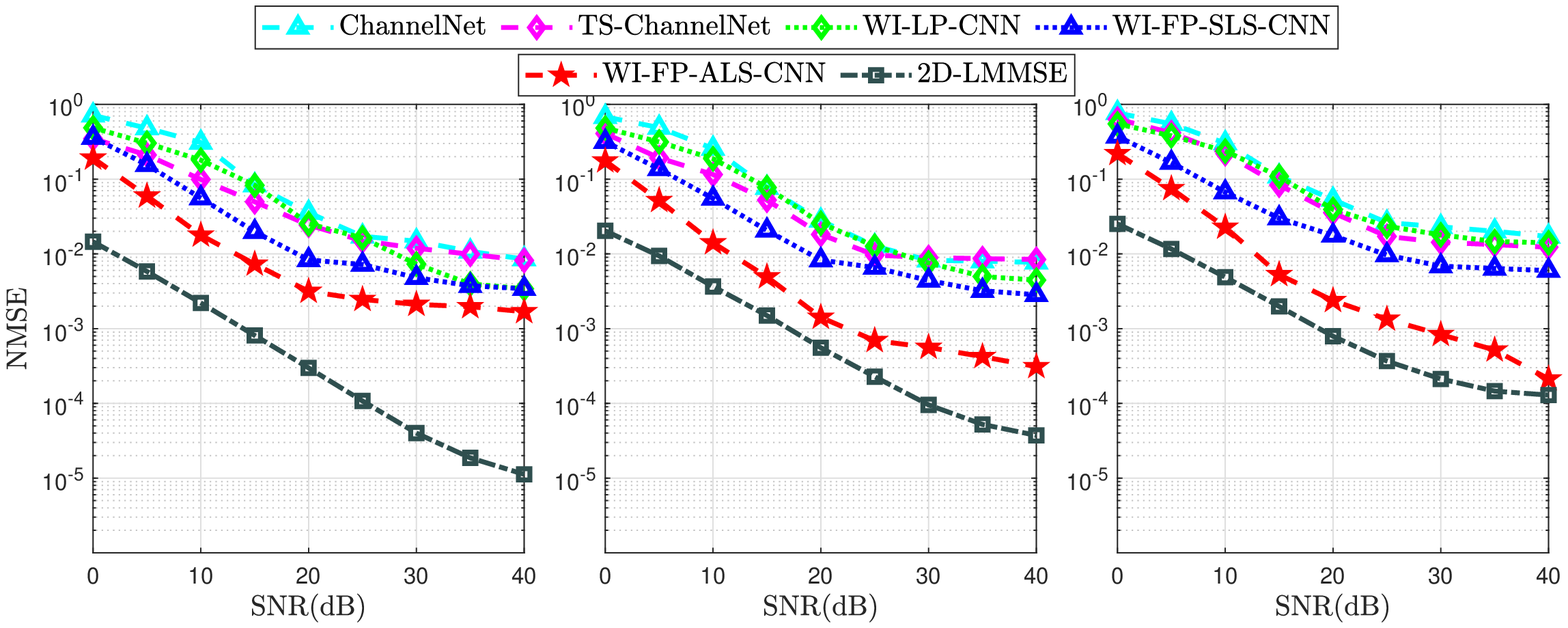} 
 	\vspace*{10pt}
	\caption{NMSE for $I=100$, mobility from left to right: low  ($v = 45~\text{Kmph}, f_{d} = 250$ Hz), high  ($v = 100~\text{Kmph}, f_{d} = 500$ Hz), very high  ($v = 200~\text{Kmph}, f_{d} = 1000$ Hz).The CNN refers to SR-CNN and DN-CNN in low and high/very high) mobility scenarios, respectively.}
	\label{fig:NMSE_DL_FBF}
\end{figure*}


\subsubsection{Modulation Order}
 Figure~\ref{fig:BER_DL_FBF} illustrates the BER performance of the studied DL-Based FBF estimators employing QPSK and 16QAM modulation orders. The 2D \ac{LMMSE} uses the channel and noise statistics in the estimation, thus leading to comparable performance in terms of the ideal case. However, the 2D-{\ac{LMMSE}} is ridden with high computational complexity. Moreover, the significant {\ac{BER}} performance superiority of the WI-CNN estimators can be observed where FP-ALS-CNN outperforms the ChannelNet as well as TS-ChannelNet estimators by at least $6$ dB and $3$ dB gain in terms of {\ac{SNR}} for a BER = $10^{-3}$. Importantly, {\ac{ChannelNet}} and {\ac{TS-ChannelNet}} estimators suffer from a considerable performance degradation that is dominant in very high mobility scenarios. This is because their performance accounts for the predefined fixed parameters in the applied interpolation scheme, where it is important to update the RBF interpolation function and the ADD-TT frequency and time averaging parameters in real-time. Furthermore, the ADD-TT interpolation employs only the previous and the current pilot subcarriers for the channel estimation at each received {\ac{OFDM}} symbol. By contrast, there are no fixed parameters in the WI-CNN estimators. The time correlation between the previous and the future pilot symbols is considered in the {\ac{WI}} interpolation matrix~{\eqref{eq:ci}}, whereas the estimated channel is considered in the overall estimation at all channel taps. These aspects lead to the superior performance of WI-CNN estimators performance, where a significant robustness is shown against high mobility with varied performance gain according to the employed pilot allocation scheme, i.e FP or LP. In addition, {WI}-CNN estimators employ  optimized {\ac{SR-CNN}} and {\ac{DN-CNN}} in accordance with the mobility condition, wherein SR-CNN is utilized in low mobility scenarios, whereas, DN-CNN is employed in high and very high mobility scenarios.}
\begin{figure}[t]
  \centering
  \includegraphics[width=1\columnwidth]{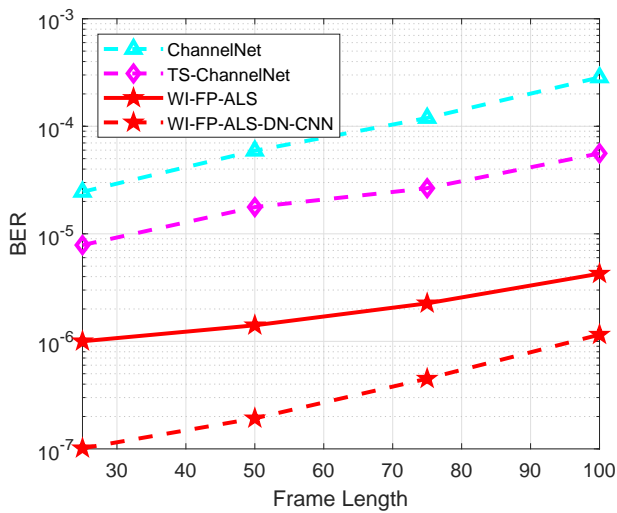}
  \caption{{BER performance of VTV-SDWW high mobility vehicular channel model employing QPSK modulation and different frame lengths.}}
  \label{fig:ber_fl}
\end{figure}

\begin{figure*}[t]
	\setlength{\abovecaptionskip}{3pt plus 3pt minus 2pt}
	\centering
	\includegraphics[width=2\columnwidth]{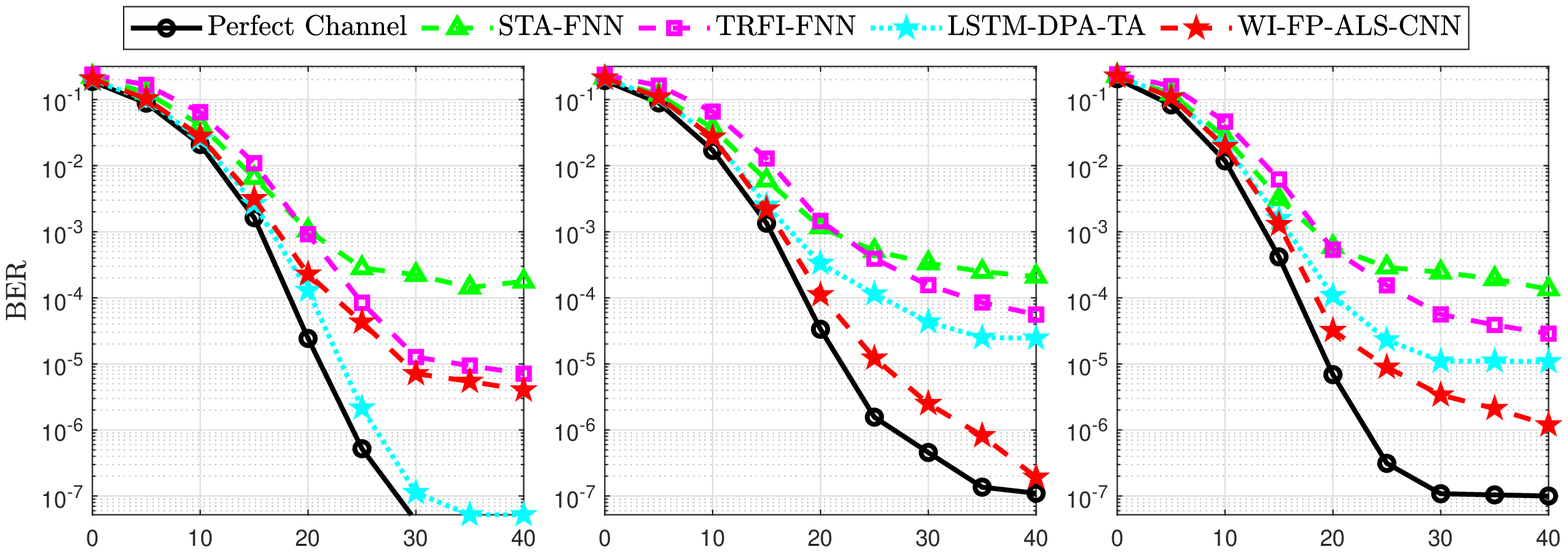}\\[-3ex]
	\subfloat[\label{BER_QPSK_DL_SBSFBF} BER performance employing QPSK.]{\hspace{.5\linewidth}} \\[-2ex]
	\vspace*{10pt}
	\includegraphics[width=2\columnwidth]{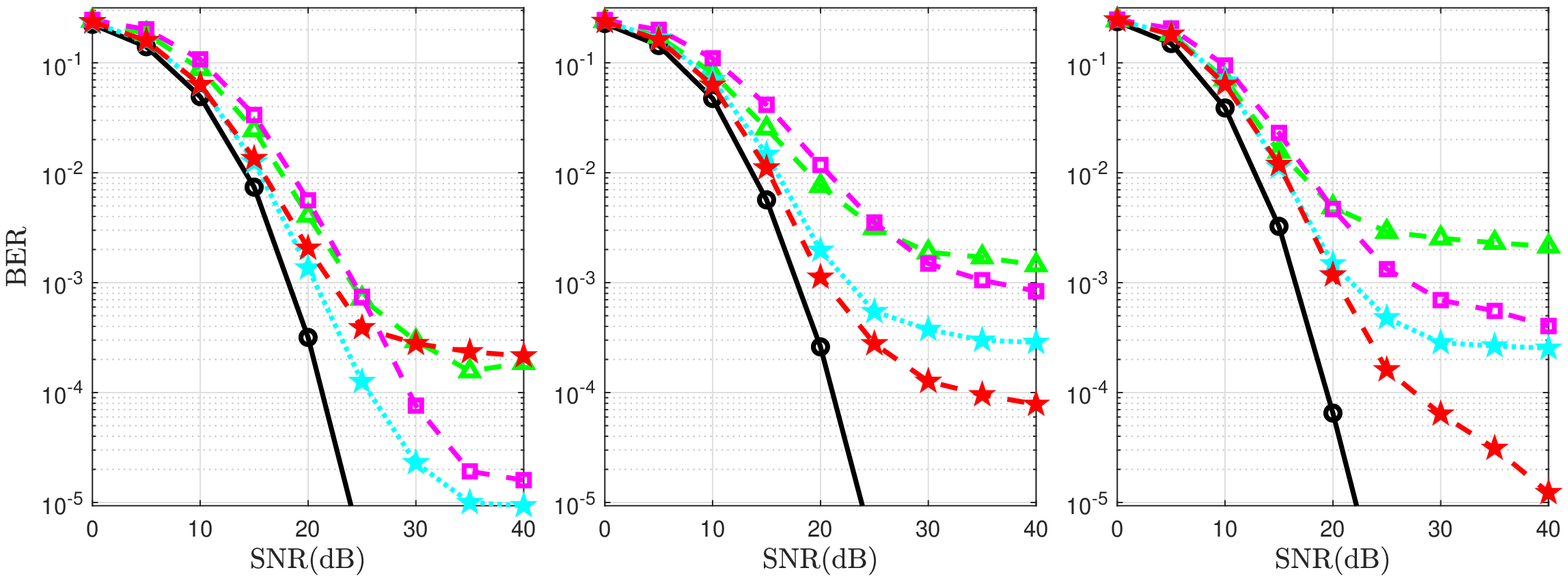}\\[-3ex]
	\subfloat[\label{BER_16QAM_DL_SBSFBF} BER performance employing 16QAM.]{\hspace{.5\linewidth}}
	\caption{BER performance employing three scenarios: (\textit{i}) first column - low mobility ($v = 45~\text{Kmph}, f_{d} = 250$ Hz) (\textit{ii}) second column - high mobility ($v = 100~\text{Kmph}, f_{d} = 500$ Hz) (\textit{iii}) third column - very high mobility ($v = 200~\text{Kmph}, f_{d} = 1000$ Hz). The CNN refers to SR-CNN and DN-CNN in low and high/very high) mobility scenarios, respectively.}
	\label{fig:BER_DL_SBSFBF}
\end{figure*}

\subsubsection{Mobility}
A degradation is observed in the overall performance of {\ac{ChannelNet}} and {\ac{TS-ChannelNet}} estimators as the mobility increases, while the WI-CNN estimators reveal a robustness against high mobility, as illustrated in \figref{fig:BER_DL_FBF}.  This is primarily attributed to the accuracy of the {\ac{WI}} interpolation, coupled with optimized {\ac{SR-CNN}} and {\ac{DN-CNN}}. Although {\ac{CNN}} processing is implemented in the {\ac{ChannelNet}} and {\ac{TS-ChannelNet}}, this post {\ac{CNN}} processing is unable to perform well due to the high estimation error of the 2D RBF and ADD-TT interpolation techniques in the initial estimation. Therefore, it can be concluded that employing robust initial estimation as the {\ac{WI}} interpolation schemes allows the {\ac{CNN}} to better learn the channel correlation with lower complexity, thereby enhancing the channel estimation.
\subsubsection{Frame Length}

\color{black}
{Figure~{\ref{fig:ber_fl}} illustrates the {\ac{BER}} performance of high mobility vehicular scenario employing QPSK modulation and different frame lengths. As can be clearly observed, the WI-FP-ALS estimator outperforms ChannelNet and TS-ChannelNet for different frame lengths without any post CNN processing. This is because of the long codeword that shows the robustness of the WI-FP-ALS estimator, unlike the 2D RBF and ADD-TT interpolation techniques that suffer from a significant estimation error even when considering a short frame. Moreover, employing the optimized DN-CNN after the WI-FP-ALS estimator significantly enhances the BER performance. 

However, although {\ac{CNN}} processing is applied in the {\ac{ChannelNet}} and {\ac{TS-ChannelNet}}, this post {\ac{CNN}} processing is unable to perform well. This is attributed to the high estimation error of the 2D RBF and ADD-TT interpolation techniques in the initial estimation. Thus, we can conclude that employing robust initial estimation as the {\ac{WI}} interpolation schemes enable the {\ac{CNN}} to better learn the channel correlation with lower complexity, thereby enhancing the channel estimation, as shown in \figref{fig:NMSE_DL_FBF}.

\begin{figure*}[t]
	\setlength{\abovecaptionskip}{3pt plus 3pt minus 2pt}
	\centering
	\includegraphics[width=2\columnwidth]{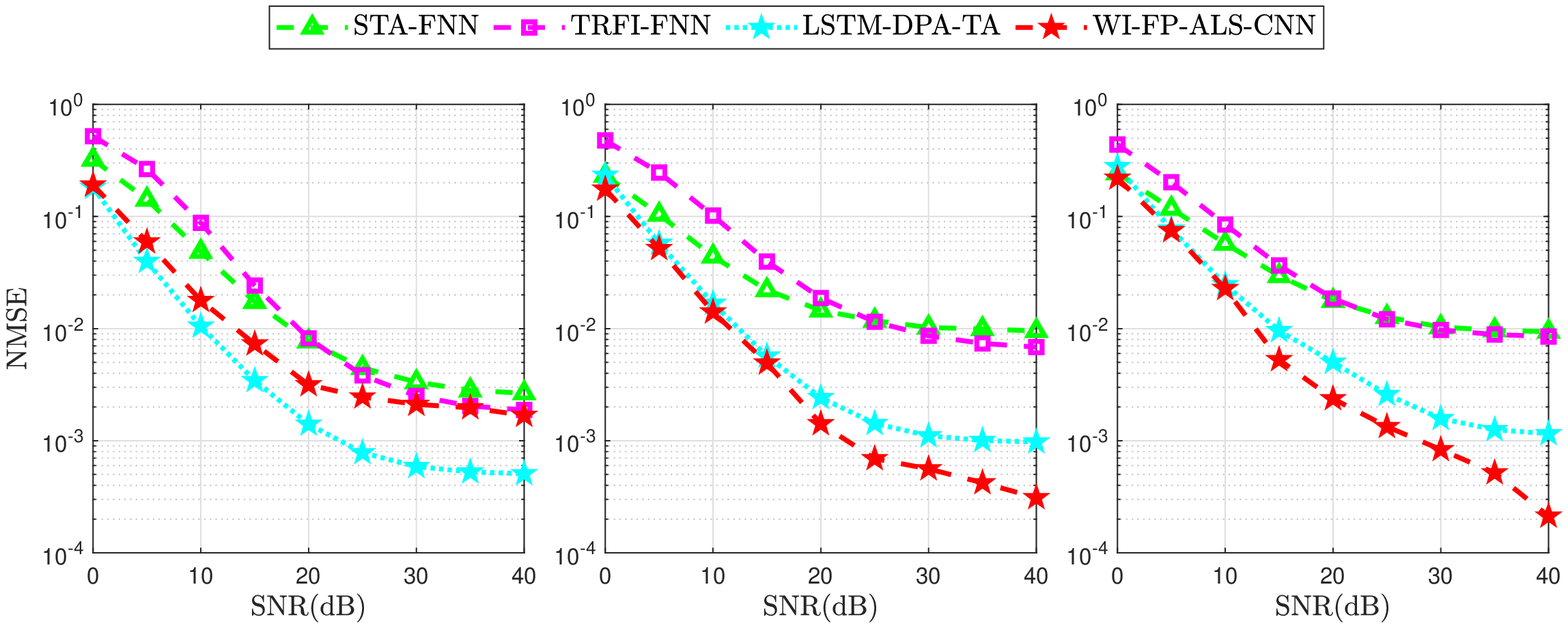}
	\vspace*{10pt}
	\caption{NMSE performance employing three scenarios: (\textit{i}) first column - low mobility ($v = 45~\text{Kmph}, f_{d} = 250$ Hz) (\textit{ii}) second column - high mobility ($v = 100~\text{Kmph}, f_{d} = 500$ Hz) (\textit{iii}) third column - very high mobility ($v = 200~\text{Kmph}, f_{d} = 1000$ Hz). The CNN refers to SR-CNN and DN-CNN in low and high/very high) mobility scenarios, respectively.}
	\label{fig:NMSE_DL_SBSFBF}
\end{figure*}
\subsection{CNN Architecture}
The {\ac{ChannelNet}} estimator employs  {\ac{SR-CNN}} and {\ac{DN-CNN}} following the 2D RBF interpolation. The employed {\ac{SR-CNN}} comprises three convolutional layers with $(v_{1} = 9 ; f_{1} = 64), (v_{2} = 1 , f_{2} = 32)$ and $(v_{3} = 5 , f_{3} = 1)$, respectively. Moreover, the {\ac{DN-CNN}} depth is $D = 18$  with $3 \times 3 \times 32$ kernels in each layer. Meanwhile, {\ac{SR-ConvLSTM}} network comprises three ConvLSTM layers of $(v_{1} = 9 ; f_{1} = 64), (v_{2} = 1 , f_{2} = 32)$ and $(v_{3} = 5 , f_{3} = 1)$, respectively, and is integrated after the ADD-TT interpolation in the {\ac{TS-ChannelNet}} estimator. The {\ac{SR-ConvLSTM}} network combines both the {\ac{CNN}} and the LSTM networks~\cite{ref_ConvLSTM_Comp}, thus increasing the overall computational complexity, as shall be discussed later. By contrast, the employed optimized SR-CNN and DN-CNN significantly reduces the complexity due to the WI estimators' accuracy. Put succinctly, the complexity of the employed CNN decreases as the accuracy of the pre-estimation increases, because low-complexity architectures can be utilized and vice versa.

\subsection{DL-Based SBS vs. DL-Based FBF estimation Scheme}

This section further examines the performance assessment of the studied estimators, where only the best DL-based SBS and FBF estimators are compared. Figures \ref{fig:BER_DL_SBSFBF} and~\ref{fig:NMSE_DL_SBSFBF} illustrate the BER and NMSE performance of the investigated {\ac{DL}}-based estimators in low, high, and very high mobility scenarios, employing QPSK and 16QAM modulation orders.

In low-mobility scenario, the LSTM-DPA-TA \ac{SBS} estimator outperforms the WI-FP-ALS-SR-CNN FBF estimator. This can be explained by the ability of LSTM to better learn the channel time correlation than the SR-CNN, since Doppler error is somehow negligible in low mobility scenario. However, in high and very high mobility scenarios, WI-FP-ALS-DN-CNN shows a significantly improved performance, outperforming the LSTM-DPA-TA SBS estimator by $3$ dB gain in terms of SNR for a BER = $10^{-4}$. In high mobility scenarios, where the Doppler error impact is high, LSTM suffers from some performance degradation as learning the time correlation between successive samples is not achievable in the low mobility scenario case. Meanwhile, DN-CNN network can significantly alleviate the impact of noise and Doppler error, where it records at least $5$ dB gain in terms of SNR for a BER = $10^{-4}$. To conclude, it can be inferred that employing LSTM network rather than FNN and DN-CNN networks leads to improved performance in low-mobility scenarios. In By contrast, DN-CNN is more useful in high as well as very high mobility scenarios because DN-CNN uses the entire pilot subcarriers within the received frame. To summarize, the time correlation between successive received OFDM symbols decreases as the mobility increases. Therefore, the performance of  LSTM suffers from performance degradation when compared with CNN. On the other hand, the CNN-based estimators become more useful than the LSTM-based estimators in high mobility scenarios.

Finally, it is observes that DL-based FBF estimators suffer from high buffering time at the receiver, because it is necessary to receive the full frame before the channel estimation begins leading to high latency. However, this buffering time is lowered in the WI-CNN estimators after dividing the received frame into sub frames so that the channel estimation process commences prior to the full frame reception. Moreover, the WI-CNN estimators also help increase the transmission data rate as fewer pilots are inserted into the transmitted frame.    

\section{Complexity Analysis} \label{complexity}

This section provides a detailed computational complexity analysis of the studied {\ac{DL}}-based {\ac{SBS}} and {\ac{FBF}} estimators. The computational complexity analysis is performed in accordance with the number of real-valued arithmetic operations,  multiplication/division and summation/subtraction   necessary to estimate the channel for one received {\ac{OFDM}} frame. Each complex-valued division requires $6$ real-valued multiplications, $2$  divisions, $2$ summations, and $1$ subtraction. In addition, each complex-valued multiplication is performed by $4$ real-valued multiplications and $3$  summations.
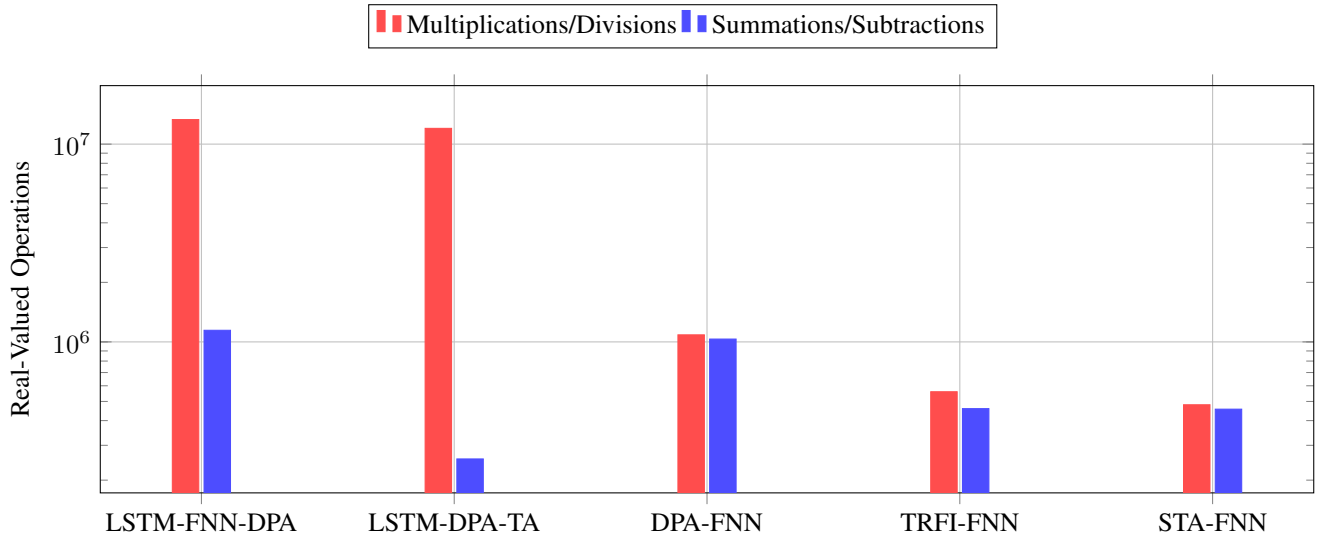
\begin {figure*}[t]
\centering
\begin{tikzpicture}
\begin{axis}[
    ybar,
    ylabel={Real-Valued Operations},
    symbolic x coords={LSTM-FNN-DPA, LSTM-DPA-TA,DPA-FNN, TRFI-FNN, STA-FNN},
    xtick=data,
    ymode=log,
    legend style={at={(0.48,+1.2)},
    anchor=north,legend columns=-1},
    nodes near coords align={vertical},
    width=2\columnwidth,
    height=7cm,
    grid=major,
    cycle list = {red!70,blue!70,red!40,black!10}
    ]
\addplot+[fill] coordinates {  (LSTM-FNN-DPA,13308800) (LSTM-DPA-TA, 12013600) (DPA-FNN,1085600) (TRFI-FNN, 559800  ) (STA-FNN, 481000 )};
\addplot+[fill,text=black!10] coordinates { (LSTM-FNN-DPA,1144800)  (LSTM-DPA-TA, 256000) (DPA-FNN,1033600)  (TRFI-FNN, 459800) (STA-FNN, 457000 )};
\legend{Multiplications/Divisions, Summations/Subtractions}
\end{axis}
\end{tikzpicture}
\caption{Computational complexity comparison of the studied DL-based SBS estimators.}
\label{fig:bar_graph_LSTM}
\end{figure*}
\begin{table*}[t]
	\renewcommand{\arraystretch}{1.5}	
	\centering
	\caption{Detailed computation complexity of the studied DL-based SBS estimators.}
	\label{tb:comp}
	\begin{adjustbox}{width=14cm}
	\begin{tabular}{|c|c|c|}
		\hline
		\textbf{Estimator} &\textbf{ Mul./Div.}  & \textbf{Sum./Sub.}   \\ \hline
		FNN($J_2$-$J_3$-$J_4$)      & $2 K_{\text{on}} J_2$ + $J_2 J_3$ + $J_3 J_4$ + $2 K_{\text{on}} J_4$  & $2 K_{\text{on}} J_2$ + $J_2 J_3$ + $J_3 J_4$ +$ 2 K_{\text{on}} J_4$ 
		\\ \hline
		LSTM ($P$)     & $P^{2} + 3P + P K_{in}$ & $4P +  K_{in} - 2$ 
		\\ \hline
		\multicolumn{3}{|c|}{Overall channel estimation}
		\\ \hline
		STA-FNN & $ 82 K_{\text{on}} + 2 K_{d} + 450$   & $70 K_{\text{on}} + 10 K_{d} + 450$      \\ \hline
		TRFI-FNN            & $94 K_{\text{on}} + 26 K_{\text{int}} + 450$  & $74 K_{\text{on}} + 30 K_{\text{int}} + 450$\\ \hline
		DPA-FNN    & $178 K_{\text{on}} + 1600$  & $168 K_{\text{on}}  + 1600$        \\ \hline
		LSTM-FNN-DPA    & $512 K_{\text{in}} + 98 K_{d} + 71040$ &  $4 K_{\text{in}} + 88 K_{d} + 6776$       \\ \hline
		LSTM-DPA-TA($64$)    & $514 K_{on} + 18 K_{d} + 16576 $  & $10 K_{on} + 8 K_{d} + 824$         \\ \hline
		LSTM-DPA-TA($128$)    & $1026 K_{on} + 18 K_{d} + 65920 $  &  $ 10 K_{on} + 8 K_{d} + 1656$       \\ \hline
	\end{tabular}
	\end{adjustbox}
\end{table*}
\subsection{DL-Based SBS Estimators}

The {\ac{DPA}} estimation implemented in the DL-based {\ac{SBS}} estimators as an initial step needs two equalization steps~{\eqref{eq: DPA_1}}, and~{\eqref{eq: DPA_2}}. Each equalization step comprises $K_{\text{on}}$ complex-valued divisions. Moreover, it needs the LS estimated channel at the preamble computed by $2 K_{\text{on}}$ summation and $2 K_{\text{on}}$ divisions. Hence, the overall computational complexity of the {\ac{DPA}} estimation is $16 K_{\text{on}}$ multiplications/divisions and $6 K_{\text{on}}$ summations/subtractions.

The {\ac{STA}} estimator applies frequency as well as time-domain averaging in addition to {\ac{DPA}}. The frequency-domain averaging~{\eqref{eq: STA_4}} coefficient is fixed ($\beta = 2$). Thus, each subcarrier requires $5$ complex-valued summations multiplied by a real-valued weight, which, in turn, are equivalent to $10$ real-valued summations, and $2$ real-valued multiplications. Consequently, the {\ac{STA}} frequency-domain averaging step requires $10 K_{d}$ real-valued summations, and $2 K_{d}$ real-valued multiplications. The {\ac{STA}} time-domain averaging step~{\eqref{eq: STA_5}} requires $4 K_{\text{on}}$ real-valued divisions, and $2 K_{\text{on}}$ real-valued summations. For this reason, the accumulated overall computational complexity of {\ac{STA}} estimator is $22 K_{\text{on}} + 2 K_{\text{d}}$ multiplications/divisions and $10 K_{\text{on}} + 10 K_{\text{d}} $ summations/subtractions.

The {\ac{TRFI}} estimator implements another two equalization steps after the {\ac{DPA}} estimation~{\eqref{eq: TRFI_1}}. Thereafter, it applies cubic interpolation as the last step. Based on the analysis performed in~\cite{ref_STA_DNN}, the computational  complexity of {\ac{TRFI}} is $34 K_{\text{on}} + 26K_{\text{int}}$ multiplications/divisions and $14 K_{\text{on}} + 30K_{\text{int}}$ summations/subtractions, where $K_{\text{int}}$ represents the number of unreliable subcarriers in each received {\ac{OFDM}} symbol.

\subsubsection{FNN-based Estimators}
For the FNN-based estimators, the DPA-FNN architecture~\cite{ref_dpa_dnn} consists of three hidden layers with $J_{1} = J_{5} = 2 K_{\text{on}}$, $J_{2} = J_{4} = 40$, and $J_{3} = 20$ neurons, respectively. Therefore, the DPA-FNN requires $4 K_{\text{on}} J_{2} + 2J_{2}J_{3}$ multiplications, and $2 K_{\text{on}}  + 2J_{2} + J_{3}$ summations.  The computational complexity of \ac{LS} and the DPA estimation are accumulated for DPA-FNN computational complexity resulting in total of $178 K_{\text{on}} +1600$ multiplications and $168 K_{\text{on}} + 1600$ summations/subtractions.
\begin{equation}
CC_{\text{FNN}} = 2 \sum_{l=0}^{L+1} {N}_{l-1} {N}_l,~ \mbox{where } {N}_0 = {N}_{L+1} = 2K_{\text{on}}.
\label{eq:DNNcomp}
\end{equation}

The STA-FNN and TRFI-FNN estimators employ a three-hidden layer FNN architecture consisting of $15$ neurons each. This FNN architecture requires $4 K_{\text{on}} J_{2} + 2J_{2}^2$, and  $2 K_{\text{on}} + 3J_{2}$ summations. This architecture is less complex when compared with the DPA-FNN one. Thus, the {\ac{STA}}-{\ac{FNN}} overall computational complexity is $ 82 K_{\text{on}} + 2 K_{d} + 450$ multiplications, and  $70 K_{\text{on}} + 10 K_{d} + 450$ summations/subtractions. Furthermore, the TRFI-FNN needs $94 K_{\text{on}} + 26 K_{\text{int}} + 450$ multiplications, and  $74 K_{\text{on}} + 30 K_{\text{int}} + 450$ summations/subtractions. The {\ac{TRFI}}-{\ac{FNN}} estimator reduces the number of multiplications as well as summations by $48\% $ and $56\%$, respectively, when compared with DPA-{\ac{FNN}}, while its computational complexity is similar to that of {\ac{STA}}-{\ac{FNN}}.
\subsubsection{LSTM-Based Estimators}
The computational complexity of the  {\ac{LSTM}} unit can be calculated with respect to the number of operations performed by its four gates. Each gate applies $P^{2} + P K_{in}$ real-valued multiplications and $3P +  K_{in} - 2$  summations apart from $3P$  multiplications, and $P$  summations required by~\eqref{eq: lstm_cell_state}, and~\eqref{eq: lstm_hidden_state}. As a result, the overall computational complexity for the {\ac{LSTM}} becomes
\begin{equation}
CC_{\text{LSTM}} = 4 (P^{2} + P K_{\text{in}} + 3P + K_{\text{in}} -2) + 4P.
\label{eq:LSTMcomp}
\end{equation}
Notably, {\ac{FNN}}-based estimators need less computation than {\ac{LSTM}}, thus achieving lower complexity. 
\begin {figure*}[t]
\centering
\begin{tikzpicture}
\begin{axis}[
    ybar,
    ylabel={Real-Valued Operations},
    symbolic x coords={ChannelNet,TS-ChannelNet,FP-ALS-DN-CNN, FP-ALS-SR-CNN},
    xtick=data,
    ymode=log,
    legend style={at={(0.48,+1.2)},
    anchor=north,legend columns=-1},
    nodes near coords align={vertical},
    width=2\columnwidth,
    height=7cm,
    grid=major,
    cycle list = {red!70,blue!70,red!40,black!10}
    ]
\addplot+[fill] coordinates {  (ChannelNet,2595149600) (TS-ChannelNet, 1180401600) (FP-ALS-DN-CNN,424235264)  (FP-ALS-SR-CNN, 35401392  )  }; 
\addplot+[fill,text=black!10] coordinates { (ChannelNet,231045600)  (TS-ChannelNet, 424060000) (FP-ALS-DN-CNN, 49764312)   (FP-ALS-SR-CNN, 5699928) }; 
\legend{Multiplications/Divisions, Summations/Subtractions}
\end{axis}
\end{tikzpicture}
\caption{Computational complexity comparison of the studied DL-based FBF estimators.}
\label{fig:bar_graph_CNN}
\end{figure*}
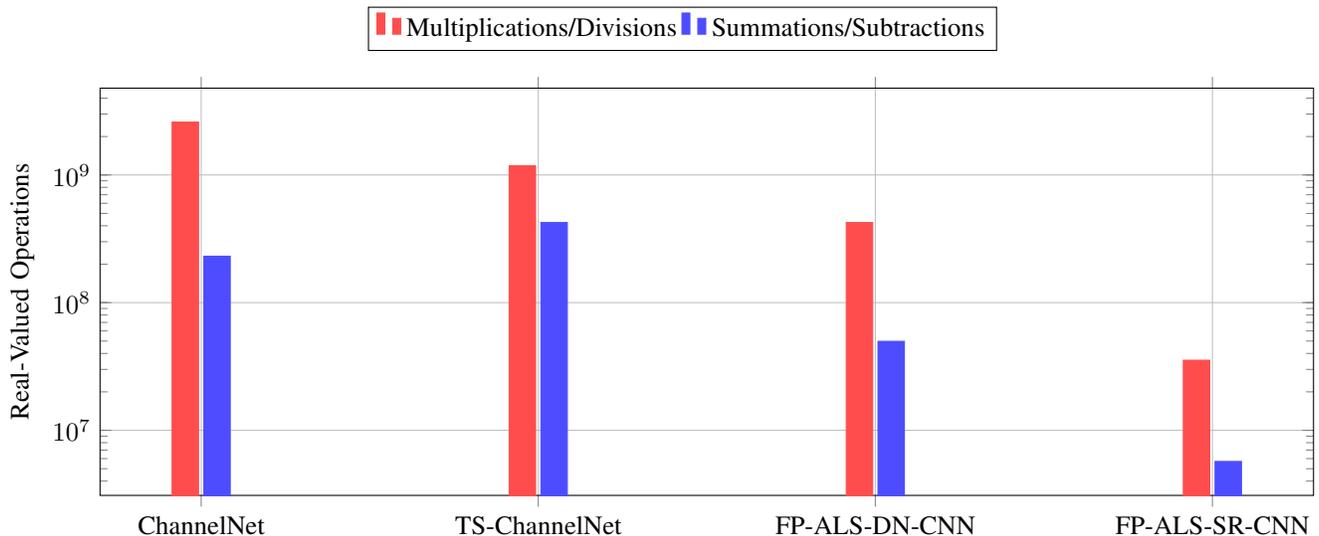
The {\ac{LSTM}}-{\ac{FNN}}-{\ac{DPA}} estimator employs one {\ac{LSTM}} unit with $P=128$ and $K_{in} = 112$, followed by one hidden layer {\ac{FNN}} network with $N_{1} = 40$ neurons. In addition, the {\ac{LSTM}}-{\ac{FNN}}-{\ac{DPA}} estimator implements the {\ac{DPA}} estimation that requires $18 K_{d}$ real-valued multiplication/division and $8 K_{d}$  summation/subtraction. Thus, the overall computational complexity of the {\ac{LSTM}}-{\ac{FNN}}-{\ac{DPA}} estimator is $512 K_{\text{in}} + 98 K_{d} + 71040$ multiplication/division and $4 K_{\text{in}} + 88 K_{d} + 6776$  summation/subtraction.

The {\ac{LSTM}}-{\ac{DPA}}-{\ac{TA}} utilizes one {\ac{LSTM}} unit with $P=128$ as {\ac{LSTM}}-{\ac{FNN}}-{\ac{DPA}} estimator. It also uses $K_{in} = 2 K_{on}$, and applies {\ac{TA}} as a noise alleviation technique to the $\hat{\bar{\ma{h}}}_{\text{LSTM-DPA}_{i,d}}$ estimated channel, that requires only $2 K_{on}$ real-valued multiplication/division and $2 K_{on}$  summation/subtraction. Hence, the {\ac{LSTM}}-{\ac{DPA}}-{\ac{TA}} estimator requires $4 P^{2} + P  (8 K_{on} + 3) + 18 K_{d} + 2 K_{on} $  real-valued multiplication/division and $13P + 10 K_{on} + 8 K_{d} - 8$  summation/subtraction.
As per this analysis, the {\ac{LSTM}}-{\ac{DPA}}-{\ac{TA}} estimator achieves less computational complexity in comparison to the {\ac{LSTM}}-{\ac{FNN}}-{\ac{DPA}} estimator. It records $9.73 \%$ and $77.63\%$ computational complexity decline in the required real-valued multiplication/division and summation/subtraction, respectively. Importantly, replacing the {\ac{FNN}} network by the {\ac{TA}} to achieve noise alleviation is the primary factor in reducing the overall computational complexity. Moreover,   the {\ac{LSTM}}-{\ac{DPA}}-{\ac{TA}} estimator outperforms the {\ac{LSTM}}-{\ac{FNN}}-{\ac{DPA}} estimator while recording lower computational complexity. As a matter of fact, employing the {\ac{LSTM}}-{\ac{DPA}}-{\ac{TA}} LSTM-based estimators as opposed to the FNN-based estimators results in $89.10\%$ and $62.18\%$ increase in the necessary multiplication/division and summation/subtraction, respectively. Nevertheless, it is possible to achieve a significant performance gain. \tabref{tb:comp} and \figref{fig:bar_graph_LSTM} reveal a detailed summary of the computational complexities for the various examined DL-based SBS estimators.
\\

\subsection{{\ac{DL}}-Based FBF Estimators}

\subsubsection{{\ac{ChannelNet}} estimator}
The {\ac{ChannelNet}} estimator utilizes the {\ac{RBF}} interpolation followed by {\ac{SR-CNN}} and {\ac{DN-CNN}} networks. Therefore, the overall computational complexity of the {\ac{ChannelNet}} estimator can be expressed as follows
\begin{equation}
\text{CC}_{{\text{ChannelNet}}} = \text{CC}_{{\text{RBF}}} + \text{CC}_{\text{SR-CNN}} + \text{CC}_{\text{DN-CNN}}.
\label{eq:CC_ChannelNet}
\end{equation}
The calculation of $\hat{\tilde{\ma{H}}}_{\text{LS}}$ requires $2 K_{p}I$ divisions. The computation of $\ma{w}_{\text{RBF}}$ requires $4 K^{2}_{p}I^{2} $ multiplications/divisions and $5 K^{2}_{p}I^{2} - 2 K_{p}I$ summations/subtractions. Meanwhile, $\hat{\tilde{\ma{H}}}_{\text{RBF}}$  requires $K_{d}I (K^{2}_{p} I^{2} + 3K_{p}I)$ multiplications/divisions and $5K_{d}K_{p} I^{2}$ subtractions/summations. Thus, the total computational complexity of the {\ac{RBF}} interpolation can be expressed by $K^{2}_{p} I^{2} ( 4 + K_{d} I ) + K_{p} I (2 + 3 K_{d} I)$ multiplications/divisions and $ K_{p} I ( 5 K_{p} I + 5 K_{d}I - 2)$  summations/subtractions.
Subsequently, the {\ac{ChannelNet}} estimator applies {\ac{SR-CNN}} followed by {\ac{DN-CNN}} in addition to the {\ac{RBF}} interpolation. $\text{CC}_{\text{SR-CNN}}$ and $\text{CC}_{\text{DN-CNN}}$ can be computed as follows
\begin{equation}
\begin{split}
\text{CC}_{\text{SR-CNN}} &= \sum_{l=1}^{\mathcal{L}} h_{l} w_{l} d_{l} v_{l}^2 f_{l}  + h_{l} w_{l} d_{l} f_{l} \\
&= \sum_{l=1}^{\mathcal{L}} h_{l} w_{l} d_{l} f_{l} (v_{l}^2 + 1).
\end{split}
\label{CC_SR_CNN}
\end{equation}
\begin{equation}
\text{CC}_{\text{DN-CNN}} = \sum_{l=1}^{\mathcal{L}}  h_{l} w_{l} d_{l} f_{l} (v_{l}^2 + 1) + \sum_{j=1}^{D} 4h_{j} w_{j} d_{j}.
\label{CC_DN_CNN}
\end{equation}

$\mathcal{L}$ signifies the number of employed {\ac{CNN}} layers. It can be noted that the second term in $\text{CC}_{\text{DN-CNN}}$ signifies the number of operations required by the batch normalization employed in the {\ac{DN-CNN}} network. Thus, the {\ac{SR-CNN}} employed in the {\ac{ChannelNet}} estimator needs $16064 K_{\text{on}} I$ multiplications/divisions as well as $4288 K_{\text{on}} I$ summations/subtractions, while the {\ac{ChannelNet}} {\ac{DN-CNN}} computations require $334080 K_{\text{on}} I$ multiplications/divisions and $38144 K_{\text{on}} I$ summations/subtractions.

\begin{table*}[t]
\renewcommand{\arraystretch}{1.5}
\centering
\caption{Detailed computation complexity of the studied CNN-based FBF estimators.}
\label{tb:comp_FBF}
\begin{adjustbox}{width=14cm}
\begin{tabular}{|c|c|c|c|c|}
\hline
\multirow{2}{*}{\textbf{Scheme}} & \multicolumn{2}{c|}{\textbf{Interpolation}} & \multicolumn{2}{c|}{\textbf{CNN}}       \\ \cline{2-5} 
                                 & \textbf{Mul./Div.}   & \textbf{Sum./Sub.}   & \textbf{Mul./Div.} & \textbf{Sum./Sub.} \\ \hline
ChannelNet  & \begin{tabular}[c]{@{}c@{}} $K^{2}_{p} I^{2} ( 4 + K_{d} I )$ \\ + $K_{p} I (2 + 3 K_{d} I)$ \end{tabular}  & \begin{tabular}[c]{@{}c@{}}$ K_{p} I ( 5 K_{p} I$  \\ + $5 K_{d}I - 2)$      \end{tabular}                &   $350144 K_{\text{on}} I$                 &      $42432 K_{\text{on}} I$              \\ \hline
TS-ChannelNet    & $24 K_{\text{on}} I + 4 L K_{\text{on}} I$     &  \begin{tabular}[c]{@{}c@{}} $18 K_{\text{on}} I$ \\ + $5 K_{\text{on}} I L$  \end{tabular}        &  $226880 K_{\text{on}} I$  &   $81472 K_{\text{on}} I$                 \\ \hline
FP-SLS-SR-CNN  &  \begin{tabular}[c]{@{}c@{}} $2 K_{\text{on}} P + 2 K_{\text{on}}$ \\ + $4 K_{\text{on}} I_{d}$ \end{tabular}       &   \begin{tabular}[c]{@{}c@{}}  $2 K_{\text{on}}$ \\ + $2 K_{\text{on}} I_{d}$ \end{tabular}                     & \multirow{4}{*}{$7008 K_{\text{on}}I_{d}$}  & \multirow{4}{*}{$1120 K_{\text{on}}I_{d}$}  \\ \cline{1-3}
FP-ALS-SR-CNN   & \begin{tabular}[c]{@{}c@{}} $4 K^{2}_{\text{on}} P + 2 K_{\text{on}} P$ \\ + $2 K_{\text{on}} + 4 K_{\text{on}} I_{d}$ \end{tabular}   &   \begin{tabular}[c]{@{}c@{}} $5 K^{2}_{\text{on}} P$ \\ + $2 K_{\text{on}} I_{d}$ \end{tabular}                     &                    &                    \\ \cline{1-3}
LP-SR-CNN    &   \begin{tabular}[c]{@{}c@{}} $2LP + 4 K_{\text{on}} L P$ \\ + $2 K_{\text{on}} + 4 K_{\text{on}} I_{d} $ \end{tabular}        & \begin{tabular}[c]{@{}c@{}} $5 K_{\text{on}} L P$ \\ + $2 K_{\text{on}} I_{d}$ \end{tabular}                         &                    &                    \\ \cline{1-3}
 \hline
FP-SLS-DN-CNN   &   \begin{tabular}[c]{@{}c@{}} $2 K_{\text{on}} P + 2 K_{\text{on}}$ \\ + $4 K_{\text{on}} I_{d}$ \end{tabular}        &     \begin{tabular}[c]{@{}c@{}} $2 K_{\text{on}}$ \\ + $2 K_{\text{on}} I_{d}$   \end{tabular}                    & \multirow{4}{*}{$84096 K_{\text{on}}I_{d}$}  & \multirow{4}{*}{$9856 K_{\text{on}}I_{d}$}  \\ \cline{1-3}
FP-ALS-DN-CNN   &   \begin{tabular}[c]{@{}c@{}}  $4 K^{2}_{\text{on}} P + 2 K_{\text{on}} P$ \\ + $2 K_{\text{on}} + 4 K_{\text{on}} I_{d}$  \end{tabular}    &   \begin{tabular}[c]{@{}c@{}} $5 K^{2}_{\text{on}} P$ \\ + $2 K_{\text{on}} I_{d}$ \end{tabular}                  &                    &                    \\ \cline{1-3}
LP-DN-CNN    &  \begin{tabular}[c]{@{}c@{}}  $2LP + 4 K_{\text{on}} L P + 2 K_{\text{on}}$ \\ + $4 K_{\text{on}} I_{d} $  \end{tabular}  & \begin{tabular}[c]{@{}c@{}} $5 K_{\text{on}} L P$ \\ + $2 K_{\text{on}} I_{d}$ \end{tabular}                       &                    &                    \\ \cline{1-3}
 \hline
\end{tabular}
\end{adjustbox}
\end{table*}

\subsubsection{{\ac{TS-ChannelNet}} estimator}
The {\ac{TS-ChannelNet}} estimator applies the \ac{ADD-TT} interpolation followed by the  {\ac{SR-ConvLSTM}} network. Hence, the overall computational complexity of the {\ac{TS-ChannelNet}} estimator can be expressed in the following manner:
\begin{equation}
\text{CC}_{{\text{TS-ChannelNet}}} = \text{CC}_{{\text{ADD-TT}}} + \text{CC}_{\text{SR-ConvLSTM}}.
\label{eq:CC_TS-ChannelNet}
\end{equation} 
The \ac{ADD-TT} interpolation first applies the {\ac{DPA}} estimation requiring $18 K_{\text{on}}$ multiplications/divisions and $8 K_{\text{on}}$ summations/subtractions. The time-domain truncation operation applied in~\eqref{eq:TT} requires $4 L K_{\text{on}}$ multiplications as well as $5K_{\text{on}} L - 2K_{\text{on}}$ summations. In the  {\ac{ADD-TT}} interpolation, the frequency-domain averaging~\eqref{eq: STA_44} requires $10 K_{\text{on}}$ summations and $2 K_{\text{on}}$ multiplications. Furthermore, the time-domain averaging step~\eqref{eq: STA_55} requires $4 K_{\text{on}}$ real valued divisions, and $2 K_{\text{on}}$ real valued summations. Thus, the overall computational complexity of the {\ac{ADD-TT}} interpolation for the whole received {\ac{OFDM}} frame requires $24 K_{\text{on}} I + 4 L K_{\text{on}} I$ real-valued multiplications/divisions, and $18 K_{\text{on}} I + 5 K_{\text{on}} I L$  summations/subtractions. The total computational complexity is expressed with respect to the overall operations implemented in the input, forget, and output gates of the {\ac{SR-ConvLSTM}} network, such that
\begin{equation}
\text{CC}_{\text{ConvLSTM}} = \sum_{l=1}^{\mathcal{L}}  h_{l} w_{l} d_{l} f_{l} (8v_{l}^2 + 30). 
\label{CC_ConvLSTM}
\end{equation}
Based on~\eqref{CC_ConvLSTM}, the {\ac{SR-ConvLSTM}} network employed in the {\ac{TS-ChannelNet}} estimator requires $226880 K_{\text{on}} I$ multiplications/divisions as well as $81472 K_{\text{on}} I$ summations/subtractions.
{\ac{TS-ChannelNet}} estimator is less complicated than the {\ac{ChannelNet}} estimator, because it employs only one {\ac{CNN}} in addition to the {\ac{ADD-TT}} interpolation, unlike the {\ac{ChannelNet}} estimator where both {\ac{SR-CNN}} and {\ac{DN-CNN}} are employed.
\subsubsection{{\ac{WI}}-CNN estimators}
The \ac{WI}-CNN estimators computational complexity primarily depends on the selected frame structure, the pilot allocation scheme, as well as the selected optimized {\ac{CNN}}. Thus, the overall computational complexity of the {\ac{WI}}-CNN estimators can be expressed as follows
\begin{equation}
\text{CC}_{{\text{WI}}} = \text{CC}_{\hat{\tilde{\ma{H}}}_{\text{WI}}} + \text{CC}_{\text{O-CNN}}.
\label{eq:CC_WI}
\end{equation}
When  full pilot symbols are inserted, two options are taken into consideration. The first option is the {\ac{SLS}} estimator, which is performed using $2 K_{\text{on}} P + 2 K_{\text{on}}$ divisions, and $2 K_{\text{on}}$ summations. The second option entails employing the {\ac{ALS}} estimator with $2  K_{\text{on}} P + 2 K_{\text{on}}$ divisions. This is followed by  $4  K^{2}_{\text{on}} P$ multiplications, and $ 5 K^{2}_{\text{on}} P$ summations. In the instance where $K_{\text{p}} = L$ pilots are inserted with each pilot symbol,  the {\ac{LS}} estimation requires $2 L P + 2 K_{\text{on}} $ divisions, $4  K_{\text{on}} L P$ multiplications, and $ 5 K_{\text{on}} L P $ summations. In a similar manner, for employing only $K_{p} = 4$ pilot subcarriers, the WI-CP estimator needs $8 P + 2 K_{\text{on}} $ divisions, $16  K_{\text{on}} P$ multiplications, as well as $ 20 K_{\text{on}} P $ summations. Following the selection of the required frame structure and pilot allocation scheme, the WI-CNN estimators apply the weighted interpolation as demonstrated in~\eqref{eq:WI_LS}.The channel estimation for each received {\ac{OFDM}} frame needs $4 K_{\text{on}}I_{d}$ divisions and  $2 K_{\text{on}}I_{d}$ summations. Finally, the optimized {\ac{SR-CNN}} is utilized in low-mobility scenario and needs $7008 K_{\text{on}}I_{d}$ multiplications/divisions and  $1120 K_{\text{on}}I_{d}$ summations/subtractions. For high-mobility scenarios, the optimized {\ac{DN-CNN}} is employed, requiring $84096 K_{\text{on}}I_{d}$ multiplications/divisions and $9856 K_{\text{on}}I_{d}$ summations/subtractions.
The {\ac{WI}}-FP-ALS records the higher computational complexity among the other WI estimators in all mobility scenarios, due to $\ma{W}_{\text{ALS}}$ calculation in~\eqref{eq:ALSP}, whereas, the {\ac{WI}}-FP-SLS estimator refers to the simplest one. 

Table~\ref{tb:comp_FBF} shows the studied estimators' overall computational complexity with respect to real valued operations. It is noteworthy that the {\ac{WI}} estimators achieve significant computational complexity decrease  in comparison to {\ac{ChannelNet}} and {\ac{TS-ChannelNet}} estimators. {\figref{fig:bar_graph_CNN}} depicts the computational complexity of the studied DL-based FBF estimators. The {\ac{ChannelNet}} and {\ac{TS-ChannelNet}} estimators are $70$ and $39$ times more complex than the FP-ALS-SR-CNN, respectively. In addition, the WI-CNN estimators achieve a minimum of $7027.35$ times less complexity than the 2D LMMSE estimator, with an acceptable {\ac{BER}} performance, which makes them a feasible alternative to the 2D LMMSE. It is also observed that FP-ALS-DN-CNN is $12$ times more complex than FP-ALS-SR-CNN since the optimized {\ac{DN-CNN}} architecture complexity employed in high and very high scenarios is higher than the optimized {\ac{SR-CNN}} architecture, which, in turn, is employed in low mobility scenarios.

\section{Conclusion} \label{conclusions}

This survey sheds light on the recently proposed {DL}-based SBS and FBF channel estimators in doubly-dispersive environments. First, we have defined the problem of signal propagation in doubly-dispersive channels. Subsequently, a review of different DL architectures employed in the doubly-dispersive channel estimation has been undertaken, followed by a detailed presentation of the studied DL-based estimators. Finally, the studied estimators have been evaluated with respect to {\ac{NMSE}}, {\ac{BER}}, and computational complexity, clearly demonstrating a significant improvement of employing DL in the channel estimation across different mobility conditions. We have shown that, while the LSTM and CNN based estimators do outperform the FNN based estimator, more computational complexity is necessary where the LSTM-based SBS estimator is $23.6$ times more complex than the FNN-based SBS estimators. Nevertheless, the complexity of the CNN-based FBF estimator exceeds the complexity of LSTM-based SBS estimator by approximately $3450$ times because of the significant difference in terms of required operations between the CNN and LSTM networks. Finally, we have observed that the choice of the channel estimator is primarily related to the applications requirements as well as affordable computational complexity. SBS estimators are more useful when the application is sensitive to latency, whereas FBF estimators can be employed if some latency can be accepted. To summarize, a trade-off between the required performance, computational complexity, and the accepted latency must first be defined to select what is the most suitable channel estimator to be employed.

\ifCLASSOPTIONcaptionsoff
  \newpage
\fi
\bibliographystyle{IEEEtran}
\bibliography{ref}
\end{document}